\documentclass[a4paper,11pt]{article}
\pdfoutput=1 

\usepackage{jheppub} 

\usepackage[T1]{fontenc} 

\usepackage{setspace}

\usepackage{physics}
\usepackage{dsfont}

\def\Tr{\text{Tr}}

\newcommand{\be}{\begin{equation}}
\newcommand{\ee}{\end{equation}}

\def\beq{\begin{eqnarray}}\def\eeq{\end{eqnarray}}
\def\be{\begin{equation}}\def\ee{\end{equation}}

\title{Aspects of  Entanglement Entropy For Gauge Theories}

 \author{Ronak M Soni,}
 \author{Sandip P. Trivedi}
 \affiliation{\it Department of Theoretical Physics,
 Tata Institute of Fundamental Research,\\  Colaba, Mumbai, 400005, India}

\vspace{5mm}

\vspace{1cm}

\emailAdd{ronak@theory.tifr.res.in}
\emailAdd{sandip@theory.tifr.res.in}

\abstract{A definition for the entanglement entropy in  a gauge theory was given recently in arXiv:1501.02593. Working on a spatial lattice, it involves embedding the physical state in an extended Hilbert space obtained by taking the tensor product of the Hilbert space of states on  each link of the lattice. This extended Hilbert space admits a tensor product decomposition by definition and allows a density matrix and entanglement entropy for the set of links of interest to be defined. Here, we continue the study of this extended Hilbert space definition with particular emphasis on the case of Non-Abelian gauge theories.

We extend the electric centre definition of Casini, Huerta and Rosabal to the Non-Abelian case  and find that it differs in an important term. We also find that the entanglement entropy does not agree with the maximum number of Bell pairs that can be extracted by the processes of entanglement distillation or dilution, and give protocols  which achieve the maximum bound. 
Finally, we compute the topological entanglement entropy which follows from the extended Hilbert space definition and show that it correctly reproduces the total quantum dimension in a class of Toric code models based on Non-Abelian discrete groups. 
}

\preprint{\parbox{3cm}{TIFR/TH/15-29}}

\begin{document} 
\maketitle
\flushbottom

\section{Introduction} \label{sec:intro}
Quantum systems are known to behave in essentially different ways from their classical counterparts. 
 Entanglement is a key  property  that characterises  quantum correlations  and   entanglement entropy provides
  a quantitative measure of some of these essential differences. For a bipartite system consisting of two parts $A, B$, with Hilbert spaces,  ${\cal H}_A, {\cal H}_B,$ respectively,  which is in a pure 
  state, $|\psi\rangle$,  the entanglement entropy $S_{EE}$
  is obtained by constructing the density matrix
  of one of the two parts, say $A$, after tracing over the second one, $B$,
  \be
  \label{dens}
  \rho_A=\Tr_{{\cal H}_B }|\psi\rangle\langle\psi|
  \ee
 and computing its von Neumann entropy
 \be
 \label{defee}
 S_{EE}=-\Tr_{{\cal H}_A}\rho_A \log\rho_A.
 \ee
 The Hilbert space of the full system in this case is given by the tensor product  
 \be
 \label{tensprod}
 {\cal H}={\cal H}_{A}\otimes {\cal H}_{B}.  
 \ee
     
      In a gauge theory the definition of the entanglement gets  more complicated. It turns out that the Hilbert space of physical states, i.e., gauge-invariant states, 
      does not admit a tensor product decomposition of the type eq.(\ref{tensprod}) in terms of the Hilbert space of states in a  region $A$   and its complement.     
        Thus, it is not clear  how to compute the entanglement described above.
      This is a very general  feature in gauge theories and it is related to the presence of non-local operators like Wilson lines, which create non-local excitations, in these theories. 
       
      Several approaches have been suggested in the literature, see, \cite{Buividovich2008b,Donnelly2011,Casini2013,Radicevic2014,Ghosh2015,Hung2015,Aoki2015,Casini2014,Donnelly2014,Donnelly2014b,Donnelly2015}, to circumvent this problem and arrive at a satisfactory definition of entanglement entropy for a gauge theory. 
      Here we will follow the approach discussed in \cite{Ghosh2015} (GST), see also \cite{Aoki2015}, which is  also related to  the earlier work, \cite{Buividovich2008b,Donnelly2011}.   In this approach, by working on a spatial lattice, a definition of entanglement entropy for a gauge theory is given as follows. One first embeds  the space of gauge-invariant states in a bigger space ${\cal H}$ obtained by taking the tensor product of the Hilbert spaces 
on each link of the lattice. This  space, ${\cal H}$,  by definition admits a tensor product decomposition in terms of the Hilbert spaces for the set of links of interest, which we denote as
${\cal H}_{in}$, and the Hilbert space for the rest of the links, ${\cal H}_{out}$. We can write,  
${\cal H}= {\cal H}_{in} \otimes {\cal H}_{out}$, which is analogous to eq.(\ref{tensprod}).  
It is then straightforward to   obtain the density matrix,
$\rho_{in}$, eq.(\ref{dens}), by taking a trace over ${\cal H}_{out}$, and from it  the entropy, eq.(\ref{defee}).
The definition given in GST works for any set of links, including in particular a set of links which are adjacent or close to each other  in space. 
We will refer to it as the extended Hilbert space definition below. 

This definition has several nice features. It is   gauge-invariant.  It meets the condition of strong subadditivity.
And it can  be shown to agree with a path integral definition of entanglement which follows from  the replica trick\footnote{
More accurately, since there can be ambiguities in the replica trick on the lattice, the extended Hilbert space definition agrees with the    path integral  implementing a particular version of   the replica trick.}. The definition applies to both Abelian and Non-Abelian theories and can be readily extended to gauge theories with matter. 

The purpose of this paper is to explore the extended Hilbert space definition further and elucidate some of its properties. 

A different  approach for defining the entanglement entropy was given in the work of Casini, Huerta and Rosabal (CHR) \cite{Casini2013}, see also \cite{Radicevic2014}. 
In this approach one works with gauge-invariant states, and the algebra of operators acting in the region $A$ and its complement $B$.
For a gauge theory it was argued that this algebra has a non-trivial centre, and this centre is responsible for  the  space of gauge-invariant states not having a  
 tensor product decomposition, eq.(\ref{tensprod}). One can overcome this obstacle by   diagonalising  the centre and going to sectors where it   takes a fixed value. 
 The Hilbert space of gauge-invariant states in each  sector then does admit a tensor product decomposition, leading to a definition of the density matrix and the entanglement entropy.  The different sectors are in fact superselection sectors since no local gauge-invariant operators, acting on either the inside or the outside, can change the sector.
 It turns out that different choices of the centre are possible and give different definitions for the entanglement on the lattice. A particular choice, called the electric centre,   corresponds to specifying the total electric flux entering the links of interest at each  vertex lying  on the boundary. By Gauss' law this is also equal, up to a sign,  to the total flux leaving the links of interest at the boundary vertices.  This choice leads to the electric centre definition of the entanglement entropy. 

 For Abelian theories it was shown  that the extended Hilbert space definition agreed with the electric centre  definition,  \cite{Casini2013,Ghosh2015}.  The extended Hilbert space definition can also  be expressed as a sum over sectors carrying different electric fluxes into the region of interest, and the contributions from each such sector agrees with what arises in the electric centre definition for the Abelian case. As mentioned above, the extended Hilbert space definition also works for   Non-Abelian theories. In this paper  we show how  the resulting entanglement entropy in the Non-Abelian case  can also be expressed as a sum   over different sectors with each  sector specifying,  in a gauge-invariant way, the electric flux entering the inside links. We then try to extend the electric centre definition, by analogy from the Abelian case,  to the Non-Abelian case,  and again obtain a result which is a  sum over the different electric flux sectors. However, interestingly, we find that the two definitions lead to different contributions in  each sector, resulting in a different final result for the entanglement entropy. 
  
  The difference between the two definitions is tied to an interesting subtlety which arises in the Non-Abelian case. 
  In both the Abelian and Non-Abelian cases the entanglement entropy gets contributions of two kinds. One kind is a sort of ``classical''  term which arises due to the system having  a  probability for  being in the different superselection sectors. 
  The second kind  is  due to  the entanglement entropy within each superselection sector. In the Non-Abelian case, in the extended Hilbert space definition, the classical term in turn has two contributions. These give rise to the first two terms in eq.(\ref{feeta1}), and eq.(\ref{feeta}), which will be discussed in section \ref{elcentre} in  more length. One contribution is analogous to the Abelian case and arises due to the probability  of being in the different  superselection sectors. The other contribution, which is intrinsically a feature of the Non-Abelian case, arises as follows.
  In sectors which carry non-trivial electric flux at a boundary vertex, the inside and outside  links which meet at this vertex by themselves  transform non-trivially under gauge transformations, but together must combine to be gauge-invariant.  Since in the Non-Abelian case representations of the symmetry group are of dimension greater than one this results in non-trivial entanglement  between the inside and outside  links and the resulting  additional  contribution to  the entropy. This contribution was also discussed in  \cite{Donnelly2011} and \cite{Hung2015,Aoki2015}. It turns out that this contribution of the classical kind is  absent in the electric centre definition.

 In quantum information theory an operational measure of entanglement entropy is provided by comparing the entanglement of a bipartite system with  a set of Bell pairs. This comparison is done by the process of entanglement distillation or dilution. It is well known that for a system with  localised degrees of freedom, like a spin system or a scalar field theory, the resulting number of Bell pairs produced in distillation or consumed in dilution, agrees with the definition given above, eq.(\ref{defee}). More correctly, this is true in the asymptotic sense, when one works with $n$ copies of the system  and takes the $n \rightarrow \infty$ limit. 
Only local operations and classical communication (LOCC) are  allowed in these processes. 

It was already mentioned in \cite{Ghosh2015}, see also \cite{Casini2013}, that for gauge theories the extended Hilbert space definition is not expected to agree with such an operational definition. In particular the total entanglement extracted in the form of Bell pairs in distillation or dilution should    be smaller. The difference arises because physical  LOCC operations can only involve  local gauge-invariant operations and these  are insufficient to extract all the entropy. 
In this paper we  also explore this question in greater depth. As was mentioned above, the entanglement entropy in the extended Hilbert space definition  gets contributions of two kinds, a classical contribution, and a quantum contribution due to entanglement within each superselection sector. We show  that in the asymptotic sense mentioned above, all the entanglement of the second kind can be extracted in distillation or dilution, but none of  the contribution of the first kind is amenable to such extraction. We make this precise by presenting protocols which achieve the maximum number of extracted Bell pairs for distillation and dilution. 
This result is not unexpected, and was anticipated in the Abelian case, \cite{Casini2013}, since local gauge-invariant operators cannot change the superselection sectors and therefore are not expected to be able to access the entropy contribution of the first kind. 

Since the full entanglement entropy in the extended Hilbert space definition cannot be operationally measured one might wonder whether the correct definition of entanglement is obtained by simply keeping that part which does agree with the measurements. It has been shown that the entanglement entropy  can be useful for characterising systems with non-trivial quantum correlations in states   with a mass gap. For example, the entanglement entropy has been used    to define  the topological entanglement entropy \cite{Kitaev2005,Levin2005}  which in turn is  related to important properties of the  system like the degeneracy of ground states on a manifold of non-trivial topology, and the total quantum dimension of excitations in the system. In section \ref{sec:degeneracy} we calculate the topological entanglement entropy for a class of  Toric code models \cite{Kitaev1997} based on non-Abelian discrete groups using the extended Hilbert space definition. We show that the correct result is obtained. This result gets contributions from both the classical and non-classical terms in general.  In the contribution from classical terms  both  the first and second terms in eq.(\ref{feeta}) which were mentioned a few paragraphs above contribute. Thus dropping these  terms,  which cannot be measured in distillation or dilution, would not  lead to the correct  answers for various physical questions tied to entanglement in  these systems.

This paper is organised as follows. We discuss the extended Hilbert space definition in section \ref{definition}, then compare to the electric centre definition which is given for  the Non-Abelian case in section \ref{elcentre}. A discussion of the entanglement  distillation and dilution protocols follows in section \ref{sec:conversion}. An analysis of the Toric code models is presented in section \ref{sec:degeneracy}. We end in section \ref{sec:strong-coupling}  with a discussion, for good measure, of  the entanglement in the ground state of an $SU(2)$ gauge theory to first non-trivial order in the strong coupling expansion. We   find that  to this order the entanglement arises solely due to the classical terms.

\section{The Definition}
\label{definition}
In this section we review the extended Hilbert space definition of entanglement entropy given in GST, see also \cite{Aoki2015}. 

\subsection{General Discussion}
\label{discussion}
We work in a lattice gauge theory in the Hamiltonian framework. The degrees of freedom  of the theory live on links $L_{ij}$ of a  spatial lattice. 
Gauge transformations $G_{V_i}$  are defined on  vertices, $V_i$. Physical states are gauge-invariant and satisfy the condition
\be
\label{condgi}
G_{V_{i}} \ket{\psi} = \ket{\psi}.
\ee
These physical states span the  Hilbert space of gauge-invariant states, denoted by ${\cal H}_{ginv}$. 

\begin{figure}[h]
   \centering
   \includegraphics[width=100mm]{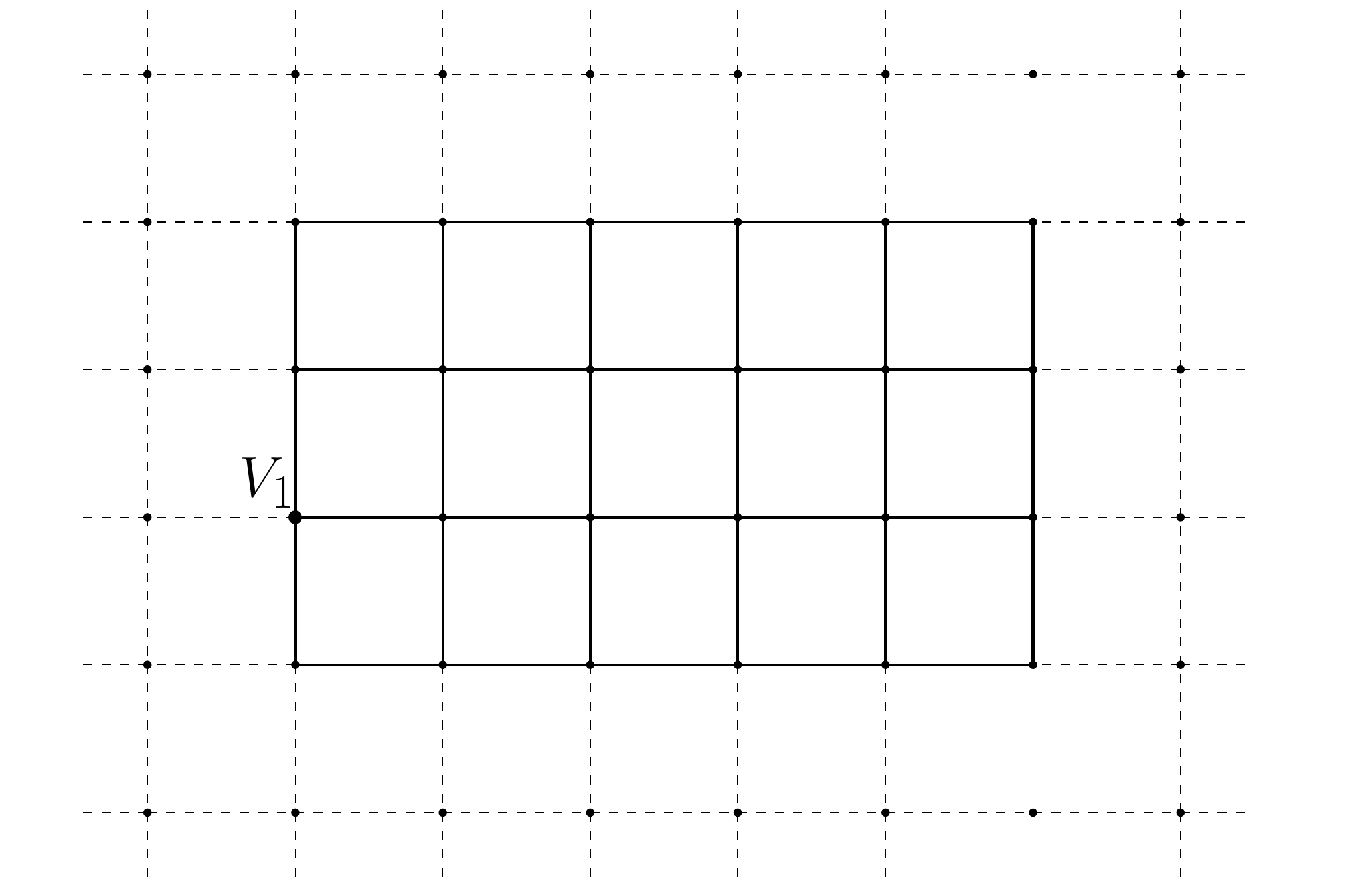}
   \caption{An example of a region in a lattice. The solid lines are the inside links, and the dotted lines are outside links. $V_{1}$ is an example of what we call a boundary vertex, one which has both inside and outside links ending on it.}
   \label{fig:generalregion}
 \end{figure}

An extended Hilbert space is  defined as follows. The degrees of freedom on each link form a Hilbert space $\mathcal{H}_{ij}$. 
The extended Hilbert space is then given by  ${\cal H} = \otimes {\cal H}_{ij}$ where the tensor product is taken over all links. 
We are interested in defining the entanglement entropy of a set of links, which we loosely call the ``inside links.'' The remaining links, not in the inside, are the ``outside links.''
In fig. \ref{fig:generalregion}, which shows a square lattice for example, the inside links are  shown as  solid lines, while the outside links are shown as dashed lines. 

${\cal H}$ admits a tensor product decomposition in terms of the Hilbert space of the links of interest, ${\cal H}_{in}$ and its complement, which is the Hilbert space of outside links, 
${\cal H}_{out}$.  
Also, if  ${\cal H}_{ginv}^{\perp}$ is the orthogonal complement of ${\cal H}_{ginv}$ then ${\cal H}$ can be written as the sum 
\be
\label{decph}
{\cal H}={\cal H}_{ginv}\oplus {\cal H}_{ginv}^\perp.
\ee

To define the entropy we regard $|\psi\rangle \in {\cal H}$, then trace over ${\cal H}_{out}$ to get the density matrix 
\be
\label{defrho}
\rho_{in}=\Tr_{{\cal H}_{out}}|\psi\rangle\langle\psi|.
\ee
The entanglement entropy is then defined as 
\be
\label{defee}
S_{EE}=-\Tr_{{\cal H}_{in}}\rho_{in} \log\rho_{in}.
\ee

This definition has several nice properties that were reviewed in GST. It is unambiguous and gauge-invariant. And it meets   the strong subadditivity condition. 
 We also note that the density matrix $\rho_{in}$ correctly gives rise to the expectation value of any gauge-invariant operator ${\hat O}$ which acts on the inside links. 
 This is because $|\psi\rangle$ is orthogonal to all  states in ${\cal H}_{ginv}^\perp$. 

 $S_{EE}$ can be expressed in  a manifestly gauge-invariant manner in terms of superselection sectors as follows. 
 We define inside vertices as those on which only links securely in the inside set end, outside vertices as those on which only outside links end, and boundary vertices as those on which some links inside and some outside links end. In   fig. \ref{fig:generalregion}, for example, $V_1$ is a boundary vertex on which three inside links and one outside link end.  A superselection sector is  specified by a vector ${\bf k}$ which gives  the total electric flux  entering the inside links at all the  boundary vertices. Labelling the boundary  vertices as $V_i, i = 1, \cdots N$, the $i^{th}$ entry $k_i$ of ${\bf k}$ is the total electric flux entering the inside at the $V_i$ boundary vertex. 
 Then the inside Hilbert space ${\cal H}_{in}$ can be written as a sum 
 \be
 \label{insk}
 {\cal H}_{in}=\oplus {\cal H}_{in}^{\bf {k}},
 \ee
 where ${\cal H}_{in}^{\bf{k}}$ is the subspace of ${\cal H}_{in}$ corresponding to the sector with flux ${\bf{k}}$. 
 
 It can be shown that 
 $\rho_{in}$ is block diagonal in the different sectors and takes the form
 \be
 \label{defrhosup}
 \rho_{in}=\oplus \rho_{in}^{\bf{k}},
 \ee
 where $\rho_{in}^{\bf{k}}$ acts on ${\cal H}_{in}^{\bf{k}}$. 
 
 $S_{EE}$ can then be expressed as 
 \be
 \label{eediff}
 S_{EE}=-\sum_{\bf{k}} \Tr_{{\cal H}_{in}^{\bf{k}}}\rho_{in}^{\bf{k}} \log\rho_{in}^{\bf{k}}.
 \ee
 Let
 \be
 \label{deftr}
 \rho_{in}^{\bf{k}}=p_{\bf{k}} {\tilde \rho}_{in}^{\bf{k}},
 \ee
 where 
 \be
 \label{trrc}
 \Tr_{{\cal H}_{in}^{\bf{k}}} \rho_{in}^{\bf{k}}=p_{\bf{k}}
 \ee
 so that 
 \be
 \label{corrt}
 \Tr_{{\cal H}_{in}^{\bf{k}}} {\tilde \rho}_{in}^{\bf{k}}=1.
 \ee
 Then eq.(\ref{eediff}) gives
 \be
 \label{defsnt2}
	 S_{EE}=-\sum_{\bf{k}} p_{\bf{k}} \log p_{\bf{k}} - \sum_{\bf{k}} p_{\bf{k}} \Tr_{{\cal H}_{in}^{\bf{k}}} {\tilde \rho}_{in}^{\bf{k}} \log {\tilde \rho}_{in}^{\bf{k}}.
 \ee
 
 The different sectors specified by the electric flux ${\bf k}$ are called superselection sectors because no gauge-invariant operator acting on the inside links alone can change these sectors.
 On the other hand gauge-invariant operators which act on both inside and outside links can change these sectors, e.g.,  Wilson loops that extend from the inside to the outside links. 
 
 The description above applies to all gauge theories without matter. These include discrete Abelian theories, like a $\mathbb{Z}_2$ theory,  continuous Abelian theories, e.g., a $U(1)$ theory, and finally Non-Abelian gauge theories, like an  $SU(2)$ gauge theory. 
 
 \subsection{Non-Abelian Theories}
 \label{nonabelian}
 For the Abelian cases a specification of the flux $k_i$ at each boundary vertex is straightforward. 
 On the other hand, for the Non-Abelian case there is an important  subtlety which we now discuss. This subtlety has to do with the fact that the irreducible representations in the Non-Abelian case are not one-dimensional. 
 
 Consider as a specific example the $SU(2)$ group, where each link, $L_{ij}$,  has as its degree 
 of freedom a $2\times 2$ matrix $U_{ij} \in SU(2)$. The corresponding state 
 in the Hilbert space ${\cal H}_{ij}$  is  denoted by $|U_{ij}\rangle$.  There are two  sets of operators ${\hat J}^a_{ij}, a= 1,2,3$, and ${\hat{\cal J}}^a_{ij}, a= 1,2,3$, which acts as generators of the algebra, acting on the left and right respectively,
 \begin{eqnarray}
 e^{i \epsilon^a {\hat J}^a_{ij}} |U_{ij}\rangle & = & \left|\left(1+i\epsilon^a {\sigma^a\over 2}\right) U_{ij}\right\rangle \label{lact} \\
 e^{i \epsilon^a {\hat{\cal J}}^a_{ij} }|U_{ij}\rangle& = & \left|U_{ij} \left(1-i\epsilon^a {\sigma^a\over 2}\right)\right\rangle.\label{ract}
 \end{eqnarray}
 The links $L_{ij}$ are oriented to emanate from vertex $V_i$ and end on $V_j$. It is easy to see that with the definitions above
 \be
 \label{relhatcal}
 {\cal {\hat J}}_{ij}^a={\hat J}_{ji}^a, a=1,2,3.
 \ee
 Also, one can show that, 
 \be
 \label{reltwoa}
 ({\hat J}_{ij})^2=({\cal {\hat J}}_{ij})^2.
 \ee
 
 The gauge transformation at $V_i$ is then generated by 
 \be
 \label{gtrans}
 G_{V^i}=e^{i \epsilon^a \sum_{j} {\hat J}^a_{ij}},
 \ee
 where the sum on the RHS is over all links emanating from $V_i$. 
 
 Now consider a boundary vertex, $V_i$.  An example is the vertex $V_1$ shown in Fig. \ref{fig:generalregion}. The total electric flux carried to outside from $V_i$ is given by 
 \be
 \label{fout}
 {\hat J}_{T,out, i }^a= \sum_{j} {\hat J}^a_{ij},
 \ee
 where the sum is over all links which are in the outside set. 
 Similarly the total flux carried inside  is given by 
 \be
 \label{fin}
 {\hat J}_{T,in,i}^a=\sum_j{\hat J}^a_{ij},
 \ee
 with the sum being over all links going in. 
 
 Gauss' law says that 
 \be
 \label{gla}
 {\hat J}_{T,out, i }^a=-{\hat J}_{T,in,i}^a.
 \ee
  The superselection sectors are  then specified by a choice of $k_i= ({\hat J}_{T,in, i})^2=({\hat J}_{T,out,i})^2$ for all boundary vertices $V_i$. 
 Since the operators, $({\hat J}_{T,in, i})^2, ({\hat J}_{T,out,i})^2$ are gauge-invariant, this is a gauge-invariant characterisation of the superselection sectors. 
 
 Now we come to the subtlety. 
 It is easy to see that  the density matrix which arises for a gauge-invariant state by tracing over ${\cal H}_{out}$ satisfies the relation
 \be
 \label{reld}
 [\rho_{in}, {\hat J}_{T, in , i}^a]=0.
 \ee
Similarly, all gauge-invariant operators acting on inside links must also commute with ${\hat J}_{T, in, i}^a$. 

Now suppose 
 \be
 \label{valki}
 k_i = j_i(j_i+1),
 \ee
 corresponding to the $2j+1$ dimensional irreducible representation of $SU(2)$. Then ${\cal H}^{\bf{k}}_{in}$ must furnish a corresponding representation of 
 ${\hat J}_{T, in, i}^a$. These two features mean that ${\cal H}^{\bf{k}}_{in}$ can be written as a tensor product
 \be
 \label{tprodh}
  {\cal H}^{\bf{k}}_{in}={\cal H}^1 \otimes {\cal H}^2\cdots \otimes {\cal H}^i \otimes \cdots  {\cal H}^n \otimes {\hat {\cal H}}^{\bf{k}}_{in},
  \ee
  where ${\cal H}^i$ is a $2j_i+1$  dimensional Hilbert space  and $i=1 \ldots n$ label the boundary vertices. All gauge-invariant operators, and also $\rho_{in}^{\bf{k}}$, act non-trivially only on the last term in the tensor product, ${\hat {\cal H}}^{\bf{k}}_{in}$,
  and are trivial on the other factors. E.g., $\rho_{in}^{\bf{k}}$ can be written as 
  \be
  \label{rhw}
  \rho_{in}^{\bf{k}}= {\bf I}_{2j_1+1} \otimes  \cdots {\bf I}_{2j_i+1} \otimes \cdots \otimes {\bf I}_{2j_n+1}  \otimes {\hat \rho}_{in}^{\bf{k}},
  \ee
  where ${\hat \rho}_{in}^{\bf{k}} $ acts on ${\hat {\cal H}}^{\bf{k}}_{in}$, and ${\bf I}_{2j_i+1}$ denotes the identity operator acting on ${\cal H}^i$. 
  A gauge-invariant operator ${\cal O}$ acting on inside links  alone cannot change the superselection sectors. In the sector $\bf{k}$ this takes the form 
  \be
  \label{giaa}
  {\cal O}^{\bf{k}}= {\bf I}_{2j_1+1} \otimes  \cdots {\bf I}_{2j_i+1} \otimes \cdots \otimes {\bf I}_{2j_n+1}  \otimes {\hat {\cal O}}^{\bf{k}},
  \ee
  where ${\hat {\cal O}}^{\bf{k}}$ acts on ${\hat {\cal H}}^{\bf{k}}_{in}$. 
  
  On the other hand, ${\hat J}_{T, in, i}^a, a=1,2,3$,  act non-trivially only on ${\cal H}^i$ and not on the other factors,
  \be
  \label{actjt}
  {\hat J}_{T, in, i}^a= {\bf I}_{2j_1+1} \otimes J^a_i \otimes \cdots \otimes {\bf I}_{2j_n+1}\otimes {\bf I},
  \ee
  where $J^a_i$ denotes a matrix  in the $2j_i+1$ dimensional representation of $SU(2)$ acting on  the ${\cal H}^i$ Hilbert space.  
  
  As a result the second term in eq.(\ref{defsnt2})  can be further split in two.  The normalised density matrix, ${\tilde \rho}_{in}^{\bf{k}}$ was defined in eq.(\ref{deftr}) , 
  eq.(\ref{corrt}),  above by rescaling the density matrix. 
  This operator can also be written in the form eq.(\ref{rhw}), 
  \be
  \label{acttwo}
  {\tilde \rho}_{in}^{\bf{k}}= {\bf I}_{2j_1+1} \otimes  \cdots {\bf I}_{2j_i+1} \otimes \cdots \otimes {\bf I}_{2j_n+1}  \otimes {\hat {\tilde  \rho}}_{in}^{\bf{k}}
  \ee
  with ${\hat{\tilde \rho}}_{in}^{\bf{k}}$ replacing
  ${\hat \rho}_{in}^{\bf{k}}$. It is now convenient to further rescale and define the operator
  \be
  \label{defre}
  {\bar \rho}_{in}^{\bf{k}}= \left( {\prod_i (2 j_i +1)} \right) {\hat{\tilde \rho}}_{in}^{\bf{k}}.
  \ee
  It is easy to see from eq.(\ref{rhw}) that this operator satisfies the condition
  \be
  \label{traa}
  \Tr_{{\hat {\cal H}}^{\bf{k}}_{in}}  {\bar \rho}_{in}^{\bf{k}}=1.
  \ee
  
  The entanglement entropy can then be written as a sum of three terms
  \be
  \label{feeta1}
  S_{EE}=-\sum_{\bf{k}} p_{\bf{k}} \log p_{\bf{k}} + \sum_{\bf{k}} p_{\bf{k}} \left(\sum_i \log(2j_i+1)\right)- \sum_{\bf{k}} p_{\bf{k}}\Tr_{{\hat {\cal H}}^{\bf{k}}_{in}}{\bar \rho}_{in}^{\bf{k}} \log {\bar\rho}_{in}^{\bf{k}} .
  \ee
  Here $\bf{k}$ denotes the different superselection sectors.  The sum over index $i$ in the second term is over all boundary vertices, with $j_i$ being the value of the total incoming angular momentum, $({\hat J}_{T, in}^a)^2$ at the $i^{th}$ boundary vertex, related to $k_i$ by eq.(\ref{valki}).
  Note that  the last trace is over the ${\hat {\cal H}}^{\bf{k}}_{in}$ subspace.

 The second term in eq.(\ref{feeta1}) then is the extra contribution which arises in the Non-Abelian case since the  irreducible representations can have dimension greater than unity. 
 Physically, the total angular momentum at a boundary vertex  must sum to zero. This results in a correlation between the in-going and out-going links at the vertex. The in-going links together give rise to a state in the  $2j_i+1$ dimension representation and so do the out-going links. 
 Then these two states combine to give a singlet of the total angular momentum emanating from the vertex.  The requirement of being in a singlet state entangles the state of the ingoing and outgoing states non-trivially, since the representations are not one dimensional. This extra entanglement results in the second term in eq.(\ref{feeta1}). 
 
 We have chosen the $SU(2)$ example for simplicity. The generalisation to any Non-Abelian group is immediate. 
 The analogue of ${\hat J}_{T, in, i }^a$, ${\hat J}_{T, out, i}^a$  are the  generators of the group acting on the ingoing and outgoing links. 
 Superselection sectors are specified by giving the irreducible representations, $r_i$, under which these generators transform, at all boundary vertices\footnote{More correctly if the outgoing links transform as the $r_i$ representation then the ingoing links must transform as the conjugate representation ${\bar r}_i$. Together these give the identity in a unique way. },  $V_i$. 
  And in the second  term in eq.(\ref{feeta1}), $\log(2j_i+1)$ is replaced by $\log d_i$ where $d_i$ is the dimension of the $r_i$ representation. As a result the entanglement is given by 
 \be
  \label{feeta}
  S_{EE}=-\sum_{\bf{k}} p_{\bf{k}} \log p_{\bf{k}} + \sum_{\bf{k}} p_{\bf{k}} \left(\sum_i \log d_i\right) - \sum_{\bf{k}} p_{\bf{k}}\Tr_{{\hat {\cal H}}^{\bf{k}}_{in}}{\bar \rho}_{in}^{\bf{k}} \log{\bar \rho}_{in}^{\bf{k}}.
  \ee 
 
 We end with some concluding comment. 
 The first term in eq.(\ref{defsnt2})  for the Abelian case is sometimes referred to as a kind of ``classical'' contribution to the entropy. It is the minimum entropy which would arise from a density matrix, eq.(\ref{defrhosup}), which meets the condition, eq.(\ref{trrc}). This minimum contribution would arise if ${\tilde \rho}^{\bf{k}}$, eq.(\ref{deftr}), in each superselection sector corresponds to a pure state. 
 Similarly,  we may regard  the first two terms in eq.(\ref{feeta}) for the non-Abelian case also  as a classical contribution. It too is the minimum contribution which arises for a density matrix which satisfies the condition, eq.(\ref{trrc}), and which has the form eq.(\ref{rhw}) required by eq.(\ref{reld}).  
 
 Gauge-invariant operators acting on only the inside or outside links cannot change the superselection sectors labelled by $\bf{k}$, and the probabilities $p_{\bf{k}}$ or the values of $d_i$. This suggests that the minimal  contribution to the entanglement  represented by the first term in eq.(\ref{defsnt2}) and the first two terms in eq.(\ref{feeta}) cannot be extracted    in dilution or distillation experiments. 
 We will see in section \ref{sec:conversion} that this expectation is indeed borne out.  
 \section{The Electric Centre Definition}
 \label{elcentre}
 
 For the Abelian case a different approach for defining the entanglement entropy using only the gauge-invariant Hilbert space of states ${\cal H}_{ginv}$ and not the extended Hilbert space was adopted in \cite{Casini2013},
 see also, \cite{Radicevic2014}. 
 
 It was shown in \cite{Casini2013} and \cite{Ghosh2015} that the electric centre choice of CHR leads to a definition of entanglement entropy that coincides with the extended Hilbert space definition given above in the Abelian case.
 Here we examine the electric centre definition  in the Non-Abelian case and find that it in fact differs from  the definition given above, precisely in the second term  on the RHS
 of eq.(\ref{feeta}).  
 
 \subsection{General Discussion}
 The essential point in the definition given by CHR is to focus on the algebra of gauge-invariant operators, ${\cal A}_{in}$, and ${\cal A}_{out}$, which act on the inside and outside links respectively. These algebras, it was argued, have a non-trivial centre,
 \be
 \label{centrea}
 {\cal A}_{in}\cap {\cal A}_{out}={\cal Z},
 \ee
 for a gauge theory, which commutes with both ${\cal A}_{in}$ and ${\cal A}_{out}$.
 This non-trivial centre is the essential obstacle to obtaining a tensor product decomposition of  ${\cal H}_{ginv}$. 
 
 To circumvent this problem, in the definition given by CHR,  one works in sectors  where the centre is diagonalised. Within each sector, labelled by say index $i$,  the intersection in eq.(\ref{centrea}) is now trivial, containing only the identity. It can then be argued that the subspace,  ${\cal H}_{ginv}^i$, of states  in this sector admits a tensor product decomposition 
 \be
 \label{tde}
 {\cal H}_{ginv}^i={\cal H}_{ginv, in}^i\otimes {\cal H}_{ginv, out}^i
 \ee
 with the operators in ${\cal A}_{in}, {\cal A}_{out}$ in the sector taking the form 
 \be
 \label{formo}
 {\cal O}^i_{in}\otimes {\bf I}^i_{out}, {\bf I}^i_{in}\otimes {\cal O'}^i_{out}
 \ee
  respectively. 
 The full Hilbert space can then be decomposed as 
 \be
 \label{fH}
 {\cal H}_{ginv}=\bigoplus_i {\cal H}_{ginv}^i= \bigoplus_i {\cal H}_{ginv, in}^i\otimes {\cal H}_{ginv, out}^i
 \ee
 and the density matrix of the inside $\rho_{in}$ can be written as a sum over components in the different sectors,
 \be
 \label{deni}
 \rho_{in}=\oplus_i \rho_{in}^i,
 \ee
 where each component $\rho_{in}^i$ is obtained by starting with ${\cal H}_{ginv}^i$ and tracing over ${\cal H}_{ginv, out}^i$. 
 
 Depending on which choice is made for the centre, one gets different definitions of the entanglement entropy.  
 As  mentioned above,  the electric centre choice of CHR leads to a definition of entanglement entropy that coincides with the extended Hilbert space definition given above in the Abelian case. The index $i$ introduced above for different sectors in this case coincides with the vector ${\bf k}$ which gives the electric flux entering the inside at each boundary vertex. 
 
 From our discussion in section \ref{discussion}, see also GST, it follows that in the Abelian case by working in the extended Hilbert space ${\cal H}$ and  considering the space of gauge-invariant states ${\cal H}_{ginv} \subset {\cal H}$, we get 
 \be
\label{wras}
{\cal H}_{ginv}=\bigoplus_{\bf{k}}  {\cal H}_{ginv, in}^{\bf{k}} \otimes {\cal H}^{\bf{k}}_{ginv,out}.
\ee
Here    ${\cal H}_{ginv,in}^{\bf{k}}\subset {\cal H}_{in}^{\bf{k}}$, consists  of all states lying in ${\cal H}_{in}^{\bf{k}}$  which are invariant under gauge transformations acting on the inside vertices. Similarly, ${\cal H}_{ginv,out}^{\bf{k}}\subset {\cal H}_{out}^{\bf{k}}$ consists of states invariant under gauge transformations acting on outside vertices.

 \subsection{Non-Abelian Case}
 \label{ecna}
  We now turn to examining the electric centre definition in the Non-Abelian case.  For concreteness we work again with the $SU(2)$ gauge theory; the generalisation to other Non-Abelian groups is quite straightforward.  
 It is clear that here too there is a non-trivial  centre in the algebra of gauge-invariant operators, ${\cal A}_{in}, {\cal A}_{out}$, which act on the inside and outside links. 
 The choice for the centre we make, which is the analogue of the electric centre choice in the Abelian case, is given by $\{{\hat J}^2_i\}$,  where this set contains the quadratic Casimirs at each boundary vertex $V_i$ \footnote{This fact was also pointed out in \cite{Radicevic2015}.}
 \be
 \label{defcas}
 {\hat J}^2_i=({\hat J}_{T, in, i})^2.
 \ee
  Physically $\{{\hat J}^2_i\}$ is a gauge-invariant way to  measure the electric flux entering a boundary vertex.
  
 That these operators lie in the centre  follows from noting that on gauge-invariant states, eq.(\ref{gla})  is valid and also that 
 $({\hat J}_{T, out, i})^2, ({\hat J}_{T,in, i})^2$ obviously commute with operators lying in ${\cal A}_{in}, {\cal A}_{out}$, respectively. Diagonalising the operators,
 eq.(\ref{defcas})  and working in sectors consisting of eigenstates with fixed eigenvalues  of the centre then leads to superselection sectors, specified by a vector $\bf{k}$, with $k_i$, eq.(\ref{valki}),  specifying the eigenvalue of ${\hat J}^2_i$ at $V_i$. 
 So far everything in the discussion is entirely analogous to what was done in the extended Hilbert space case; it also ties in with the general discussion of the CHR approach in the beginning of this section.

 On general grounds, mentioned at the beginning of the section, one expects that the states of ${\cal H}_{ginv}$ lying in each of these sectors should now admit  a tensor product decomposition of the form eq.(\ref{tde}), with the label $i$ being identified with ${\bf{k}}$. The gauge-invariant  operators  restricted to this sector must also take the form, eq.(\ref{formo}). 
 
 To compare  the electric centre definition we are developing here and the extended Hilbert space description  above we now relate
  the   spaces ${\cal H}_{ginv, in}^{\bf{k}}$, ${\cal H}_{ginv,out}^{\bf{k}}$,  with those arising  in  the extended Hilbert space description discussed in section \ref{nonabelian}.
 
 Notice that the  extended Hilbert space  ${\cal H}$ can be written as 
 \be
 \label{tprodext}
 {\cal H}=\bigoplus {\cal H}^{\bf{k}},
 \ee
 where 
 \be
 \label{prod2}
 {\cal H}^{\bf{k}}={\cal H}_{in}^{\bf{k}}\otimes {\cal H}_{out}^{\bf{k}}.
 \ee
 
 Just as ${\cal H}_{in}^{\bf{k}}$ can be expanded as eq.(\ref{tprodh}) so also  ${\cal H}_{out}^{\bf{k}}$ can be written as,
 \be
 \label{exphout}
 {\cal H}_{out}^{\bf{k}}={\cal H}^{'1}\otimes \cdots \otimes {\cal H}^{'i} \otimes \cdots \otimes {\cal H}^{'n}\otimes {\hat {\cal H}}^{\bf{k}}_{out}.
 \ee
 The dimensions of ${\cal H}^{i}$, eq.(\ref{tprodh}) and ${\cal H}^{'i}$, eq.(\ref{exphout})
 are the same for $i=1, \cdots n$, and equal to  $2j_i+1$. 
 
 There is a subspace ${\hat{\cal H}}_{ ginv,in}^{\bf{k}}$ of ${\hat {\cal H}}_{in}^{\bf{k}} $ which is gauge-invariant with respect to all 
 gauge transformations acting on the inside vertices,
 and similarly ${\hat {\cal H}}_{ ginv,out}^{\bf{k}}\subset {\hat {\cal H}}_{out}^{\bf{k}}$ which is gauge-invariant with respect to all gauge transformations on 
 the outside vertices. 
 
 We now argue that the spaces ${\cal H}^{\bf{k}}_{ginv, in}, {\cal H}^{\bf{k}}_{ginv, out}$,  can be identified with 
 ${\hat{\cal H}}_{ginv,in}^{\bf{k}}, {\hat {\cal H}}_{ ginv,out}^{\bf{k}}$,  respectively. 
 
 To show this we consider a gauge-invariant state  $|\psi\rangle$ which lies in the sector ${\bf{k}}$. This state can be embedded in the  ${\cal H}^{\bf{k}}$
 sector of the extended Hilbert space. 
 In fact, since the state is invariant with respect to gauge transformations acting on the inside and outside vertices it lies  in the subspace
 \be
 \label{subex}
 {\cal H}^{'1}\otimes \cdots  \otimes {\cal H}^{'n}\otimes {\hat {\cal H}}^{\bf{k}}_{ginv,out} \otimes {\cal H}^{1}\otimes \cdots  \otimes {\cal H}^{n}\otimes 
 {\hat {\cal H}}^{\bf{k}}_{ ginv,in}.
 \ee
 
 Gauge-invariance at the boundary vertices requires that the components in  ${\cal H}^1, {\cal H}^{'1}$, and similarly for all the other boundary vertices,
 pair up to be singlets under the boundary gauge transformations. This means that any state $|\psi\rangle$ must be of the form
 \begin{eqnarray}
  |\psi\rangle&=&\left(\sum_{\alpha_1,\alpha_1'}c_{\alpha_1,\alpha_1'}|\alpha_1\rangle\otimes |\alpha_1'\rangle \right) \otimes \left(\sum_{\alpha_2,\alpha_2'}c_{\alpha_2,\alpha_2'}|\alpha_2\rangle\otimes |\alpha_2'\rangle\right)\cdots  \left(\sum_{\alpha_n,\alpha_n'}c_{\alpha_n,\alpha_n'}|\alpha_n\rangle\otimes |\alpha_n'\rangle \right) \nonumber \\
 &&\otimes \sum_{A,b}C_{Ab} |A\rangle_{in}\otimes |b\rangle_{out} \label{depf}.
 \end{eqnarray}
 Here $|\alpha_1\rangle, |\alpha_1'\rangle$, with $ \alpha_1, \alpha_1'=1,\cdots,  2j_1+1$ are a basis for ${\cal H}^1, {\cal H}^{'1}$ respectively and so on.
 The coefficients $c_{\alpha_1,\alpha_1'}$ are such that the state is a singlet with respect to gauge transformations at the vertex $V_1$
 and similarly for the other boundary vertices. In the last term  $\{|A\rangle\}, \{|b\rangle\}$ are bases  in ${\hat {\cal H}}^{\bf{k}}_{ginv,in}, {\hat {\cal H}}^{\bf{k}}_{ ginv,out}$,
 respectively, so this term  gives a state in ${\hat {\cal H}}^{\bf{k}}_{ ginv,in}\otimes {\hat {\cal H}}^{\bf{k}}_{ ginv,out}$.
 Note that for all  states  in this sector  the first $n$ terms in the tensor product above remain the same and  only the last term which involves the state in ${\hat {\cal H}}^{\bf{k}}_{ginv,in} \otimes {\hat {\cal H}}^{\bf{k}}_{ginv,out}$ changes. 
 
 This shows that there is a one to one correspondence between states in the space ${\cal H}_{ginv}^{\bf{k}}$, which arose in the electric centre discussion  and states in ${\hat {\cal H}}^{\bf{k}}_{ginv,in} \otimes {\hat {\cal H}}^{\bf{k}}_{ginv,out}$, which we defined using the extended Hilbert space.  The map is one-to-one because given any state in ${\hat {\cal H}}^{\bf{k}}_{ginv,in} \otimes {\hat {\cal H}}^{\bf{k}}_{ginv,out}$, we can obtain a state in ${\cal H}_{ginv}^{\bf{k}}$ by appending the first $n$ terms which ensure gauge-invariance with respect to boundary gauge transformations. Moreover from the form of a gauge-invariant operator ${\cal O}_{in}^{\bf{k}}$, eq.(\ref{giaa})  which acts on inside links  and a similar form for operators acting on the outside links we see that acting on ${\hat {\cal H}}^{\bf{k}}_{ginv,in} \otimes {\hat {\cal H}}^{\bf{k}}_{ginv,out}$ these operators  take the form, 
 ${\hat {\cal O}}_{in}^{\bf{k}}\otimes {\bf I}_{out}$, ${\bf I}_{in}\otimes {\hat {\cal O}}^{\bf{k}}_{out}$ respectively in ${\hat {\cal H}}^{\bf{k}}_{ginv,in} \otimes {\hat {\cal H}}^{\bf{k}}_{ginv,out}$.
 
 It then follows, from the general arguments at the beginning of the section, that ${\cal H}^{\bf{k}}_{ginv,in}, {\cal H}^{\bf{k}}_{ginv,out}$, which we defined in the electric centre discussion above  can be identified with 
 ${\hat {\cal H}}^{\bf{k}}_{ginv,in}, {\hat {\cal H}}^{\bf{k}}_{ginv,out}$, respectively,
 \begin{eqnarray}
 {\cal H}^{\bf{k}}_{ginv,in}&\simeq& {\hat {\cal H}}^{\bf{k}}_{ginv,in},\label{ida} \\
 {\cal H}^{\bf{k}}_{ginv,out} & \simeq& {\hat {\cal H}}^{\bf{k}}_{ginv,out}.\label{idb} 
 \end{eqnarray}
 We remind the reader that ${\hat {\cal H}}^{\bf{k}}_{ginv,in}, {\hat {\cal H}}^{\bf{k}}_{ginv,out}$   were  defined  by working in  the extended Hilbert space, ${\cal H}$. 
 
 As a result, it follows from eq.(\ref{fH}) and eq.(\ref{ida}), eq.(\ref{idb}), that   the Hilbert space of gauge-invariant states can be expressed as a sum
 \be
 \label{ginvs}
 {\cal H}_{ginv}= \bigoplus {\hat {\cal H}}^{\bf{k}}_{ginv,in}\otimes {\hat {\cal H}}^{\bf{k}}_{ginv,out}.
 \ee
 
 \subsection{Entanglement Entropy}
 We  now come to the main point of this section, namely the  difference between the entanglement entropy which arises from the electric centre description given above  and the extended Hilbert space definition of the entanglement entropy. 
 
 To calculate the difference we need to relate the density matrix $\rho_{in}$ in the two cases. 
 The density matrix in the superselection sector $\bf{k}$ for the extended Hilbert space is given in eq.(\ref{rhw}). For a gauge-invariant state $|\psi\rangle$, 
 it is easy to see that ${\hat \rho}_{in}^{\bf{k}}$ has support only on  ${\hat {\cal H}}^{\bf{k}}_{ginv,in}\subset {\hat {\cal H}}^{\bf{k}}_{in}$.  Up to a normalisation, we now argue that the density matrix in a sector $\mathbf{k}$ in the electric centre definition is in fact ${\hat \rho}_{in}^{\bf{k}}$ restricted to ${\hat {\cal H}}^{\bf{k}}_{ginv,in}$.  By construction the expectation value for any gauge-invariant operator can be calculated by  working in the extended Hilbert space.  For an operator acting on  the inside links we get
   \be
 \label{correxp}
 \langle{\cal O}_{in}\rangle= \sum_{\bf{k}} \Tr_{ {\cal H}_{in}^{\bf{k} } }  {\cal O}_{in}^{\bf{k}} \rho_{in}^{\bf{k}}.
 \ee
 From eq.(\ref{rhw}), eq.(\ref{giaa}),   we see that 
 \be
 \label{condaa}
 \Tr_{{\cal H}_{in}^{\bf{k}}}  {\cal O}_{in}^{\bf{k}} \rho_{in}^{\bf{k}}=\left(\prod_i(2j_i+1) \right)\Tr_{{\hat {\cal H}}_{ginv, in}^{\bf{k}}} {\hat {\cal O}}_{in}^{\bf{k}} \hat{\rho}^{\bf{k}}_{in},
\ee
 where we have used the fact that the  trace on the RHS can be restricted to ${{\hat {\cal H}}_{ginv, in}^{\bf{k}}}$, since ${\cal O}$ is gauge-invariant and $\rho^{\bf{k}}_{in}$ has  support only in  ${{\hat {\cal H}}_{ginv, in}^{\bf{k}}}$. From eq.(\ref{deftr}), eq.(\ref{trrc}), eq.(\ref{defre})  it then follows that the density matrix in sector ${\bf{k}}$ in the electric centre definition is given by
  \be
 \label{defehoe}
 \rho_{in, ec}^{\bf{k}}= p_{\bf{k}}{\bar \rho}_{in}^{\bf{k}}
 \ee
 From eq.(\ref{traa})  and the fact that ${\bar \rho}_{in}^{\bf{k}}$ has support on only ${{\hat {\cal H}}_{ginv, in}^{\bf{k}}}$ it follows that 
 \be
 \label{norm}
 \Tr_{{\hat {\cal H}}_{ginv, in}^{\bf{k}}} {\bar \rho}_{in}^{\bf{k}}=1
 \ee
 
 It then follows that the entanglement entropy which  in the electric centre definition is given by 
 \be
 \label{entec}
 S_{EE,ec}=-\sum_{\bf{k}} \Tr_{{\hat {\cal H}}_{ginv, in}^{\bf{k}}} \rho_{in,ec}^{\bf{k}} \log\rho_{in,ec}^{\bf{k}},
 \ee
 which becomes 
 \be
 \label{entcc}
 S_{EE, ec}=-\sum_{\bf{k}} p_{\bf{k}} \log p_{\bf{k}} -\sum_{\bf{k}} p_{\bf{k}} \Tr_{{\hat {\cal H}}_{ginv, in}^{\bf{k}}} {\bar \rho}_{in}^{\bf{k}} \log{\bar \rho}_{in}^{\bf{k}}.
 \ee
 Comparing  we see that the second term in eq.(\ref{feeta1})  is missing. More generally, the second term in eq.(\ref{feeta}) will be missing,  as was mentioned at the start of this section.\footnote{The trace in the last term in eq.(\ref{feeta1}), eq.(\ref{feeta}) can be restricted to ${\hat{\cal H}}_{ginv,in}^{\bf{k}}$ since 
 ${\bar \rho}_{in}^{\bf{k}}$ has support only on this subspace of ${\hat{\cal H}}_{in}^{\bf{k}}$.}
 The physical origin of this missing term was discussed in section \ref{nonabelian}. It arises in the extended Hilbert space definition because the total angular momentum carried by inside and outside links meeting at a boundary vertex must sum to zero, giving rise to non-zero entanglement between the inside and outside links. 
 In the electric centre definition  we work  with the Hilbert space of gauge-invariant states ${\cal H}_{ginv}$ which can be decomposed only in terms of 
 ${\hat{\cal H}}_{ginv,in}^{\bf{k}}, {\hat{\cal H}}_{ginv,out}^{\bf{k}}$, eq.(\ref{ginvs}). As a result this additional contribution is absent.

\section{Entanglement Distillation and Dilution} \label{sec:conversion}
It is a standard practice to compare the entanglement in a bipartite system with a reference system typically taken to be a set of  Bell pairs. The comparison is done using the processes of entanglement dilution and distillation
which involve only Local Operations and Classical Communication (LOCC), see \cite{Bennett1996,PreskillTimeless,NielsenBook,WildeBook}. 
For a system with local degrees of freedom, like a spin system, where the Hilbert space of states admits a tensor product decomposition, it is well known that the number of Bell pairs which are consumed in dilution or produced in distillation,  is given by the entanglement entropy, 
\be
\label{EE1a}
S_{EE}=-\Tr(\rho \log(\rho)).
\ee
More accurately, this statement is true in the asymptotic limit where we take $n$ copies of the system and find that the number of pairs used up in dilution, $k'$, and those produced in distillation, $k$,  both tend to the limit,
\be
\label{lima}
\lim_{n\to \infty} {k'\over n} = \lim_{n\to \infty}{k\over n} = S_{EE}.
\ee
where $k'$ approaches the limit from above and $k$ from below.
The starting point  for distillation is that Alice and Bob share $n$ copies of the state $\ket{\psi}$ and each has $k$ unentangled reference qubits. 
Distillation is the process of converting the qubits into $k$ Bell pairs by using up the entanglement in $\ket{\psi}$.
The set-up for dilution is that Alice and Bob share $k'$ Bell pairs and Alice has  $n$ copies of the bipartite system AC in the state $|\psi\rangle$  at her disposal. 
By LOCC operations the state of the system $C$ is then teleported to Bob, using the Bell pairs, resulting in  the $n$ copies of  state $|\psi\rangle$ now being shared between Alice and Bob. The minimum number of Bell pairs required for this is $k'$.
We then  find that eq.(\ref{lima}) holds (these statements can be made more precise using $\epsilon$'s and $\delta$'s and the analysis below will be done in this more careful manner). 

In contrast, it was argued in CHR  and GST that for the extended Hilbert space definition of entanglement entropy eq.(\ref{lima})  is no longer true for  
gauge theories. This is because of the presence of superselection sectors which  cannot be changed by local  gauge-invariant operations, as reviewed in section \ref{definition}.  The existence of superselection sectors  imposes an essential limitation on extracting the full entanglement entropy using dilution or distillation. 
In the Abelian case the entanglement entropy   can be written as a sum of two terms given in eq.(\ref{defsnt2}).  Similarly in the Non-Abelian case it can be written as the sum of three terms given in eq.(\ref{feeta}). 
One expects that the  first term  in eq.(\ref{defsnt2})  in the Abelian case, and  the first two terms  in eq.(\ref{feeta})  for the Non-Abelian case cannot be extracted using local  gauge-invariant operations.

In this section we will  carry out a more detailed investigation and find that this expectation is indeed borne out. For distillation we will find that the state shared between Alice and Bob at the end is still partially entangled with a residual entanglement entropy  that cannot be reduced any further. This entropy is indeed given by the first term in eq.(\ref{defsnt2})  for the Abelian case and the sum of the first and second terms  of eq. (\ref{feeta})  in the Non-Abelian case.  In dilution we will find that Alice and 
Bob need to start with a partially entangled state  so that at the end of the dilution process they  can  share $n$ copies of the state $|\psi\rangle$.  The entanglement in this starting state
in dilution, per copy, will be the same as the residual entanglement left after distillation. 
The number of Bell pairs extracted in distillation and consumed in dilution in this way will turn out to be equal asymptotically,
\be
\label{limb}
\lim_{N\to \infty} {k'\over N} = \lim_{N\to \infty}{k\over N}
\ee
But this limit  will not equal $S_{EE}$. Instead, the limit, i.e., maximum number of Bell pairs which can be  extracted, will equal  the second term in eq.(\ref{defsnt2})  for the Abelian case, and the third term in eq.(\ref{feeta}) in the Non-Abelian case.

A similar question  relating entanglement to dilution and distillation  for a simpler superselection rule, total number conservation, was answered by Schuch, Verstraete and Cirac (SVC) in \cite{Schuch2003,Schuch2004}.
In that case, local operations cannot change the total number of particles on either Alice or Bob's side.
If a state has 1 particle on each side, then $n$ copies of the state can be transformed to another state with $n$ particles on each side; that is, there is still just one superselection rule.
In our case, however, each copy has superselection sectors arising from the fact that the Gauss law constraint must be met  in every copy. 
As a result, with $n$ copies we get $n$ sets of superselection sectors, one for each copy. 



\subsection{A Toy Model}  \label{toymodel}

We will find it convenient to carry out the analysis in a toy model consisting of just $4$ qubits. It will become clear as we proceed that our  conclusion can be easily  generalised also to  Abelian and Non-Abelian gauge theories. 
In the discussion below, at the risk of introducing somewhat confusing conventions, we label a basis for  each qubit, as  $\ket{1}$ and $\ket{0}$, with $\sigma^3 \ket{1}=\ket{1}, \sigma^3\ket{0}=-\ket{0}$. 

We will regard this $4$ qubit system, as a bipartite system, with the first two qubits being the first part, $A$, and the third and fourth qubits being the second part, $B$. 
In order to model the constraint of gauge-invariance we will impose a condition on this system. Namely, that the only allowed states are  those for which the second and third qubits take the same value. Thus an allowed or physical state in our toy model has the form
\be
\label{fqs}
\ket{\psi}=\sum_{i,j,k=0}^{1} \sqrt{p_j} a_{ijk} \ket{i,j}_A\ket{j,k}_B =\sum_j \sqrt{p_j} \ket{\psi_j}_{AB},
\ee

where 
\be
\label{defpj}
|\psi_j\rangle_{AB}=\sum_{i,k=0}^1 a_{ijk} \ket{i,j}_{A} \ket{j,k}_B
\ee
and we have normalised $|\psi_j\rangle$ so that 
\be
\label{npj}
\braket{\psi_{j}}{\psi_{j}} = 1.
\ee
The constraint that the middle two qubits  in eq.(\ref{fqs}) are  the same is the analogue of Gauss' law, which in  a gauge theory without matter 
 implies that  the electric flux leaving a region is the same as that entering its complement.

The allowed operators in this toy model are those which preserve the constraint and therefore  change the middle  two qubits together so that they continue to be equal. This is analogous the fact that the allowed operators in a gauge theory  are gauge-invariant. 
As a result,  the full algebra of allowed  operators is generated by
\begin{equation}
  \mathcal{G} = \left\{ \sigma^{z}_{1}, \sigma^{x}_{1}, \sigma^{z}_{2}, \sigma^{x}_{2} \otimes \sigma^{x}_{3}, \sigma^{z}_{3}, \sigma^{z}_{4}, \sigma^{x}_{4} \right\}.
  \label{eqn:full-algebra-four-qubits}
\end{equation}
Here  we are using the  notation that the superscript denotes the kind of Pauli matrices, $\sigma^x, \sigma^y, \sigma^z$, 
while  the subscript indicates which qubit the matrices  act on.

There are two subalgebras, $\mathcal{A}$ and $\mathcal{B}$, which act on $A$ and $B$, of allowed sets of operators,  generated by
\be
\label{subala}
{\cal G}_{\mathcal{A}} = \{\sigma^{z}_{1}, \sigma^{x}_{1},   \sigma^{z}_{2}\},
\ee and 
\be
\label{subalb}
{\cal G}_{\mathcal{B}} = \{\sigma^{z}_{3}, \sigma^{z}_{4},  \sigma^{x}_{4} \}, 
\ee
respectively. 
The centre of each subalgebra is generated by $\{\sigma^{z}_{2} = \sigma^{z}_{3} \}$. The operators in ${\cal A}$ and ${\cal B}$ are the analogue of gauge-invariant operators which act entirely in the inside or outside links in the gauge theory. The operator $\sigma^{x}_{2} \otimes \sigma^{x}_{3}$ is the analogue of a Wilson loop which crosses from one region to its complement changing the electric flux. 

In the discussion below, in an abuse of terminology, to emphasise the analogy  we will sometimes refer to the allowed operators in the algebra generated by ${\mathcal {G}}$  as gauge-invariant operators and 
to the operators in the subalgebras ${\cal A}, {\cal B}$ as local gauge-invariant operators. 

There are two superselection sectors in this toy model  given by  two values of the middle two qubits, $j=1,0$ in eq.(\ref{fqs}). It is clear that the subalgebras ${\cal A}, {\cal B}$, 
acting on $A$ or $B$ cannot change these superselection sectors. 
An important comment worth noting    is that if we consider $n$ copies of $|\psi\rangle$, each copy will have its own superselection sectors, resulting in  a total of $2^n$ superselection sectors.  

Let us end with a few more important  comments. Starting from $|\psi\rangle$ in eq.(\ref{fqs}) and tracing over $B$ we get a  density matrix
\be
\label{aal}
\rho=\left[
\begin{array}{rr}
\rho_0 & 0 \\
0 & \rho_1
\end{array}
\right]
\ee
 for the state in $A$. $\rho$ is block diagonal with the   two blocks referring to   the two superselection sectors where the second qubit takes values $0,1$ respectively. Each block is a $2\times 2$ matrix.   $\rho_0, \rho_1$ are the unnormalised  density matrices in these two sectors  with 
 \begin{eqnarray}
  \Tr(\rho_0)&  =  & p_0, \label{normrho0} \\
  \Tr(\rho_1) & = & p_1, \label{normrho1} 
  \end{eqnarray}
  being the probabilities for the two superselection sectors.

  Now consider an allowed local unitary   operator ${\hat U} $.   Under the action of this operator
  \be
  \label{actrho}
  \rho \rightarrow {\hat U} \rho {\hat U}^{\dagger}.
  \ee
  Since $U$   cannot  change the superselection sector  it  is  therefore also block diagonal. 
  We  then conclude that any such operator leaves $p_0,p_1$ unchanged. 
  Similarly, consider a generalised measurement. This  corresponds  to a set of operators, $\{M_i\}, i=1, \cdots N$ with 
  \be
  \label{condme}
 \sum_i M_i^\dagger M_i= {\bf 1}.
  \ee
  Under it the density matrix transforms as
  \be
  \label{denstr}
  \rho \rightarrow \sum_i M_i \rho M_i^\dagger.
  \ee
       An allowed measurement must be block diagonal in the basis above since it cannot change the superselection sector (i.e. each $M_i$ must be block diagonal). 
       It then follows from eq.(\ref{denstr}) that this also leaves $p_0, p_1$ unchanged, as discussed in SVC.

  Next, let us  write the state $\ket{\psi}$, eq.(\ref{fqs})  in its Schmidt basis in every sector,
\begin{equation}
  \ket{\psi_{j}} = \sum_{I=0}^{1}\sqrt{\alpha_{I,j}} \ket{I^{j},j}_{A} \ket{j,I^{j}}_B,
  \label{schmidt}
\end{equation}
Note that the $I$ label takes two values, and has a superscript when it appears as a label for states because the Schmidt  basis for the first and fourth qubit is independent in every superselection  sector. E.g.,    $\ket{0^{0}} \ne \ket{0^{1}}$.

It is easy to see that the Von Neumann entropy of the density matrix for $A$ in this state is 
\be
\label{enttoy}
S= S_c + S_q, \ee
were the  first term is,
\be
\label{t1}
S_c=-p_0 \log p_0 - p_1 \log p_1,
\ee
and the second term is 
\be
\label{defha}
S_q=- [ p_0\{\alpha_{1,0} \log\alpha_{1,0}+\alpha_{0,0}\log\alpha_{0,0} \}] -[p_1\{\alpha_{1,1}\log\alpha_{1,1} + \alpha_{0,1} \log \alpha_{0,1}\}].
\ee

These terms have the following interpretation. $S_c$ is the entropy which arises due to the probability of being in different superselection sectors. In fact, it  is the minimum entropy which can arise from any density matrix, eq.(\ref{aal}), subject to the constraint 
eq.(\ref{normrho0}), eq.(\ref{normrho1}),
which we have seen is preserved by any allowed unitary operator or measurement.
$S_q$ on the other hand 
is the average of the entanglement entropies in the two superselection sectors weighted by the probability of being in these sectors. 

The distillation and  dilution protocols we will discuss next involve local gauge-invariant operators. Since, as noted above, these  do not change $p_0$, $p_1,$ we do not 
expect  these protocols  to  be able to extract  the entropy in the first term, $S_c$. Instead,  an efficient protocol should at most to be able to extract the second term, $S_q$. 
We will see that the protocols we discuss  below indeed meet this expectation.  

We had mentioned towards the end of section \ref{definition}  that there are classical type of contributions which  arise in the entanglement of gauge theories, corresponding to the first term in eq.(\ref{defsnt2}) and the first two terms in eq.(\ref{feeta}). The $S_c$ term, eq.(\ref{t1}), in the toy model we are considering here is in fact analogous to those contributions \footnote{Actually, as we will see later,  for  the Non-Abelian case $S_c$ is analogous to the first term in eq.(\ref{feeta}) which is also the classical type of contribution we get in the electric centre definition, eq.(\ref{entcc}).}  eq.(\ref{feeta}).  We will return to the connection with gauge theories again at the end of this section  after analysing  the toy model in more detail.

\subsection{The Distillation Protocol}\label{ssec:distillation}

We start by taking $n$ copies of the state denoted by $\ket{\psi}^{\otimes n}$ with Alice having access to the $n$ copies in $A$ and Bob to the $n$ copies in $B$. In the discussion below the different copies will be labelled by the index $r=1, 2, \cdots n$. 
Next, we   consider a fuzzy measurement that Alice performs on her $n$ copies of the $A$ subsystem, corresponding to the  operator
\be
\label{op1}
{\hat O}_1= \sum_{r=1}^n \left({1+ \sigma^{z,r}_2 \over 2}\right).
\ee
Here, we remind the reader that  the label $z$ denotes the $\sigma^z$ Pauli matrix, the subscript $2$ indicates that it acts on the $2^{\rm nd}$ qubit in eq.(\ref{fqs}) 
 and the label $r$ specifies which copy among the $n$ the operator acts on.  
If the result of this measurement is $m_1$, then  the state after the measurement is given by the symmetric superposition,
\be
\label{saftera}
|\psi_1\rangle={1\over \sqrt{N_1}}\sum_{(j_r)|m_1}|\psi_{j_r}\rangle_A\otimes |\psi_{j_r}\rangle_B,
\ee
where $\sum_{(j_r)|m_1}$ indicates that the sum is over all lists $(j_r)$, subject to the constraint that  $\sum_r j_r = m_1$. The normalisation $N_1$ in eq.(\ref{saftera}) takes the value 
\be
\label{defnaa}
N_1={N \choose m_1}
\ee
 and is the total number of terms in the symmetric sum. 

Next, Alice  measures two additional variables, 
\be
\label{op2}
{\hat O}_2=\sum_{r=1}^n (|1^1\rangle_r \langle1^1|_r ) \left({1+ \sigma^{z,r}_2 \over 2}\right)
\ee
and 
\be
\label{op3}
{\hat O}_3= \sum_{r=1}^n (|1^0\rangle_r\langle1^0|_r ) \left({1-\sigma^{z,r}_2\over 2}\right).
\ee

Here $|1^1\rangle_r\langle1^1|_r$ stands for a projector which acts on the first qubit of the $r^{\rm th} $ copy, with 
 $|1^1\rangle$ denoting the state  of the first qubit in the Schmidt basis for the superselection sector  with $j=1$, see discussion after eq.(\ref{schmidt}) above. 
Similarly $|1^0\rangle_r\langle1^0|_r$ also stands for a projector of the $r^{th}$ copy, this time to a Schmidt basis state in the $j=0$ superselection sector 

If these two measurements give rise to the results $s_1$ and $t_1$ respectively then the state after these measurements is a symmetric superposition 
of $N_{T_{1}}$ terms 
\be
\label{fsysup}
|\psi'\rangle= {1\over \sqrt{N_{T_1}}} \sum_{(j_r, I_r)} \prod_r |I^{j_r}_r, j_r\rangle _A\otimes |j_r, I_{r}^{j_{r}}\rangle_B.
\ee
were the sum is over all lists $(j_r, I_r) $ with the following properties:
\begin{eqnarray}
\sum j_r & =  & m_1 \label{sumjr} \\
\sum\nolimits^{'} I_r & =  & s_1\label{sumi1} \\
\sum\nolimits^{''} I_r &= & t_1. \label{sumi2}
\end{eqnarray}
Here $\sum'$ indicates a  restricted sum over $r$ where only those terms with $j_r=1$ contribute, while $\sum^{''}$ indicates a restricted sum where only those values of $r$ with $j_r=0$ contribute. Note that 
$N_{T_1}$  in eq.(\ref{fsysup}) is given by 
\be
\label{defnt}
N_{T_1}= {n \choose m_1} {m_1 \choose s_1} {n-m_1 \choose t_1}.
\ee

Now we come to an important point. The $N_T$ terms in eq.(\ref{fsysup}) arise as follows: 
There are ${n \choose m}$ distinct values  that the set $\{j_1, j_2, \cdots j_n\}$ can take subject to the constraint that $\sum_rj_r=m$. 
Each of these choices corresponds to a different superselection sector. Working with only local gauge-invariant operators we do not expect  to extract the 
entanglement between the qubits Alice and Bob share due to this choice of superselection sectors. 
For each choice of $\{j_1, j_2, \cdots j_n\}$ and each possible value of $s$ and $t$ we have ${m \choose s} {n-m \choose t}$ different states corresponding to different choices of the $I$ labels. We would expect to be able to extract the entanglement in these states while preserving the superselection sectors for each copy. Below we see in a precise sense that this expectation is indeed met. 

To begin,  consider the case where 
\be
\label{valp}
2^p={m_1\choose s_1} {n-m_1\choose t_1},
\ee
for some integer $p$. 
In this case we provide both Alice and Bob  with $p$ qubits which are unentangled and in the state 
\be
\label{refsa}
|0\rangle^{\otimes p}_A\otimes |0\rangle^{\otimes p}_B,
\ee
 where A and B  refer to the $p$ qubits in Alice and Bob's possession respectively.  
A unitary transformation can now be carried out in each superselection sector, i.e. in each sector labelled by a distinct choice of the set $\{j_1, j_2 \cdots j_n\}$, by Alice  which swaps the  component of $|\psi'\rangle$, eq.(\ref{fsysup}),  in this superselection sector, which she has access to, 
 with that of the $p$ qubits.  Bob carries out a similar swap on his side as well. Since the state in eq.(\ref{fsysup})  is a symmetric superposition with the same number of summands, $2^p$, in each superselection sector this results in maximally entangling the two sets of $p$ qubits. The   resulting state of the full system consisting of the original $n$ copies and the additional qubits is then given by 
\be
\label{ressa}
\left({1\over \sqrt{N_1}}\sum_{(j_r)|m_1} \prod_r |0^{j_{r}},j_r\rangle_A \otimes |j_r,0^{j_{r}}\rangle_B \right)\bigotimes |\Phi_+\rangle^{\otimes p},
\ee
where 
\be
\label{defphi}
|\Phi_+\rangle=(|00\rangle+|11)/\sqrt{2}
\ee
refers to a Bell pair shared between Alice and Bob. 

We emphasise that the components of $|\psi'\rangle$, eq.(\ref{fsysup}),  corresponding to different superselection sectors, have the same number of terms in a symmetric superposition. This ensures that the above swap leaves the  reference qubits unentangled with the state of the system.


Let us also note that a swap transformation, which we will also often use below, is a unitary transformation defined as follows.  Let ${\cal H}$ and ${\tilde {\cal H}}$ be two Hilbert spaces of the same dimension and $|\phi\rangle \otimes |\chi\rangle\in {\cal H}\otimes {\tilde {\cal H}}$ , then under a swap transformation $S$ which takes ${\cal H}\otimes {\tilde{\cal H}} \rightarrow {\cal H}\otimes {\tilde{\cal H}} $,
\be
\label{defswap}
|\phi\rangle\otimes |\chi\rangle\rightarrow |\chi\rangle\otimes |\phi\rangle.
\ee

In the limit when we start with a large number of copies, $n \gg 1$, it is easy to see that  with probability $1-\epsilon$ 
\be
\label{rn1}
2^{n S_q ( 1-\delta)} <  {m_1\choose s_1} {n-m_1\choose t_1}< 2^{n S_q  (1+ \delta)},
\ee
where $S_{q}$ is the entropy eq.(\ref{defha}),
and $\delta$ is small, 
\be
\label{magd}
\delta \sim {1 \over \sqrt{n}}.
\ee
We also remind the reader that in eq.(\ref{defha})  $\alpha_{I,j}$ are  the coefficients which arise in the expansion in the Schmidt basis, eq.(\ref{schmidt}), 
and $p_0, p_1$ are the probabilities for the superselection sectors $j=0,1$ respectively,  eq.(\ref{fqs}). 

Eq.(\ref{rn1}) arises by noting that for $n\gg 1$ the expectation value  for $m_1$ is 
\be
\label{expm1}
\langle m_1\rangle=n p_1
\ee
and for $s_1, t_1$ are 
\be
\label{exps1t1}
\langle s_1\rangle=np_1\alpha_{1,1}
\ee
and 
\be
\label{expt1}
\langle t_1\rangle=np_0 \alpha_{1,0}
\ee
respectively. And by noting that fluctuations about this mean are suppressed by a fractional power of order $1/\sqrt{n}$. 
We also note that $p$ defined in eq.(\ref{valp})  satisfies the condition
\be
\label{condpa}
p>n S_q(1-\delta),
\ee
which follows from eq.(\ref{rn1}). 

Of course, in general $p$ defined by eq.(\ref{valp}) will not be an integer. In this case we can repeat this whole procedure multiple times. 
Each time  starting  with $n$ copies of the state $|\psi\rangle$ Alice  measures operators ${\hat O}_1, {\hat O}_2, {\hat O}_3$, obtaining in the $l^{th}$ iteration values $m_l, s_l, t_l$ respectively. 
After $L$ total iterations the total number of copies is 
\be
\label{totalcopy}
n_T= n L
\ee
 and  the resulting state will  consist of a symmetric superposition of several superselection sectors. The number of terms in each superselection sector now will be 
\be
\label{totalf}
N_T={m_1\choose s_1} {n-m_1 \choose t_1} {m_2\choose s_2} {n-m_2 \choose t_2} \cdots {m_L \choose s_L} {n-m_L \choose t_L}.
\ee
For $L$ sufficiently large one can in this way ensure that 
\be
\label{ens}
2^{k}\le N_T \le 2^k(1+\epsilon)
\ee 
for some integer $k$. 
Since $2^k(1+\epsilon) <2^{k+1}$, to proceed Alice and Bob can then take $k+1$ reference qubits each,  which are initially  totally unentangled in a state analogous to eq.(\ref{refsa}),  and swap their  state with that of the system in any given superselection sector.  More accurately, Alice carries out this swap between the  reference qubits and the component of the state of the system she has access to and similarly for Bob. 
 A measurement can then be made projecting the state of the qubits to a $2^k$ dimensional subspace with probability $1-\epsilon$. 
This will lead to $k$ extracted Bell pairs which are in a totally entangled state. 

It follows from eq. (\ref{rn1}), eq.(\ref{defha}) that the ratio 
\be
\label{finalreasa}
\lim_{n_T \rightarrow \infty} {k\over n_T}= S_q
\ee
This distillation procedure will work with probability $1-\epsilon$, as mentioned above. 
With a large enough number of copies,  $\epsilon$ can be made arbitrarily small. 
 
Note that in eq.(\ref{ressa}) the  two parts of the n copies of the original system  still  share some residual  entanglement, since the choice of superselection sectors in both parts is correlated.
This is also true more generally if we repeat the procedure $L$ times. The residual entanglement  is a consequence of the superselection sectors, i.e. of gauge-invariance. 
Local gauge-invariant operations cannot change the superselection sectors and thus acting with only these operators does not allow  this remaining entanglement to be extracted. 
The final state of the original system, after $L$ iterations is 
\begin{eqnarray}
  |\psi^{fin}\rangle=&&\left({1\over \sqrt{N_1}}\sum_{(j_{r_1})|m_1}\prod_{r_1}|0^{j_{r_1}},j_{r_1}\rangle_A\otimes |j_{r_1},0^{j_{r_1}}\rangle_B\right) \bigotimes \cdots \bigotimes 
\left({1\over \sqrt{N_l}}\sum_{(j_{r_l})|m_l}\prod_{r_l}|0^{j_{r_l}},j_{r_l}\rangle_A \otimes  |j_{r_l},0^{j_{r_l}}\rangle_B\right) \nonumber \\
&&\bigotimes \cdots \bigotimes \left({1\over \sqrt{N_L}}\sum_{(j_{r_L})|m_L}\prod_{r_L}|0^{j_{r_L}},j_{r_L}\rangle_A\otimes |j_{r_L},0^{j_{r_L}}\rangle_B\right), \label{psifinal}
\end{eqnarray}
 where
 \be
 \label{valnl}
 N_l={n \choose m_l}.
 \ee
\subsection{The Dilution Protocol}
In the usual discussion of dilution one considers a situation where  at the start Alice and Bob share a large number of Bell pairs and also  Alice has at her disposal $n$ copies of the state $|\psi\rangle$.
The state $|\psi\rangle$ is a bipartite system whose two parts, both in possession of Alice, we refer to as A' and C. 
Local operations are now carried out on the n copies of  C by Alice and local operations are also carried out by Bob in a coordinated way made possible by classical communication. 
We then ask what is the minimum number of  Bell pairs that are  needed so that  after the protocol is over the entanglement between A' and C is  now shared  between Alice and Bob. 
More precisely, what is the minimum number of Bell pairs needed,   as $n \rightarrow \infty$, so that Alice and Bob share the  $n$ copies of $|\psi\rangle$ with very high fidelity. 

The new feature in our discussion is that we are dealing with a system with superselection sectors. 
Local allowed operators cannot change these superselection sectors. Thus acting with only these operators we cannot hope to extract all the entanglement entropy eq.(\ref{enttoy}) between $A'$ and $C$ and transfer it to lie between Alice and  Bob.

In order to construct the state $|\psi\rangle$ with high fidelity shared between Alice and Bob it will therefore be necessary to start with a partially entangled state between Alice and Bob. This state will carry the ``barebones'' entanglement 
 given by the first term eq.(\ref{t1}) which we do not expect to be able to transfer. 
 Then we will show that the rest of the entanglement, eq.(\ref{defha}), in the asymptotic limit of a large number of copies, $n$, with high fidelity can be obtained by dilution of a set of shared Bell pairs.
 It will turn out that the minimum number of Bell pairs required in this process  $k'$ indeed meets the condition
  \be
 \label{vaptp}
\lim_{n \rightarrow \infty} {k'\over n} = \lim_{n_T \rightarrow \infty} {k\over n_T}= S_q,
 \ee
 where $k, n_T$ are defined near eq.(\ref{finalreasa}) and $S_q$ is defined in eq.(\ref{defha}). 
 
 The  partially entangled state we will  start with in the protocol is given by 
 \be
 \label{strpsi}
 |\psi^{in}\rangle={1\over \sqrt{D}} \sum_{m=\bar{m}(1-\delta)}^{m=\bar{m}(1+\delta)} \sum_{(j_r)|m} \prod_r  \sqrt{p_r}|0^{j_r},j_r\rangle_A\otimes |j_r,0^{j_r}\rangle_B.
  \ee
 Here,  $D$ is the overall normalisation required  to make this state of unit norm, 
 \be
 \label{valma}
 {\bar m}=n p_1
 \ee
  is the expectation value of $\sum j_r $ in the $n \rightarrow \infty$ limit,  and 
  \be
  \label{valdelta}
  \delta \sim O(1/\sqrt{n}). 
  \ee
 The state in eq.(\ref{strpsi}) is a bipartite state whose two parts denoted by $A$ and $B$  are accessible to Alice and Bob respectively.
 
 The rest of the steps involved in the protocol are similar to the standard discussion of dilution. 
 We start with $n$ copies of $|\psi\rangle$  whose two parts we refer to as $A'$ and $C$ both of which are in possession of Alice. 
 From eq.(\ref{fqs})  we see that 
 \be
 \label{wpsin}
 |\psi\rangle^{\otimes n}= \sum_{(j_r)}\sum_{(I_r)}\prod_r \sqrt{p_r}\sqrt{\alpha_{I_r,j_r}}|I^{j_r}_r, j_r\rangle_{A'} \otimes |j_{r}, I_{r}^{j_r} \rangle_C.
 \ee
 Here $\sum_{(j_r)}$ stands for the sum over all lists $\{j_1, j_2, \cdots j_n\}$, and $\sum_{(I_r)}$ for the sum over all lists  $\{I_1, I_2 \cdots I_n\}$.  
 
 Next,  Alice   Schumacher compresses the state $|\psi\rangle^{\otimes n}$ by projecting onto a typical subspace. This can be done by measuring suitable fuzzy operators. 
 The state is then give by 
 \begin{eqnarray}
  |\Phi\rangle={\cal N} \sum_{m=\bar{m}(1-\delta)}^{m=\bar{m}(1+\delta)}\sqrt{p_1^{m}}\sqrt{p_0^{n-m} }
 \sum_{s=\bar{s}(1-\delta_2)}^{s=\bar{s}(1+\delta_2)}&&\sum_{t=\bar{t}(1-\delta_3)}^{t=\bar{t}(1+\delta_3)}\sqrt{\alpha_{1,1}^s}\sqrt{\alpha_{0,1}^{m-s}} \sqrt{\alpha_{1,0}^{t}}\sqrt{\alpha_{0,0}^{n-m-t}}  \nonumber\\
 &&\sum_{(j_r, I_r)}\prod_r 
 |I_r^{j_r}, j_r\rangle_{A'} \otimes |j_{r}, I_{r}^{j_r}\rangle_C, \label{wfproj}
 \end{eqnarray}
 where ${\cal N}$ is an overall normalisation to make $|\Phi\rangle$ of unit norm, and  ${\bar m}$ and $\delta$ are given in eq.(\ref{valma}) and eq.(\ref{valdelta}) respectively.  Also note that 
 $\bar{s}, \bar{t}$ are dependent on $m$ and  given by 
 \begin{eqnarray}
 \bar{s} & = & m \alpha_{1,1} \label{valbars}\\
 \bar{t} & = & (n-m)\alpha_{1,0}. \label{valbart}
 \end{eqnarray}
 They are the expectation values for the operator ${\hat O}_2, {\hat O}_3$, eq.(\ref{op2}), eq.(\ref{op3}), respectively for 
 \be
 \label{valmb}
 \sum_r j_r=m.
 \ee
 Note also that we take $\delta_2, \delta_3\sim O(1/\sqrt{n})$ in eq.(\ref{wfproj}). And  finally  that $\sum_{(j_r,I_r)}$ denotes a sum over lists $\{(j_1, I_1), (j_2, I_2), \cdots (j_n,I_n)\}$ such that  $\sum_r j_r=m$;  
  $\sum'I_r=s$, and $\sum^{"} I_r=t$, where $\sum'$ denotes a  sum in which only terms   with $j_r=1$ contributes, and  $\sum^{"}$ denotes a  sum in 
  which only  terms with  $j_r=0$  contribute. 
  
 For any $\epsilon>0$ by adjusting $\delta, \delta_2,\delta_3$ we can get a state $|\Phi\rangle$ which approximates $|\psi\rangle^{\otimes n}$ with fidelity $1-\epsilon$.

 To proceed we first consider a particular   superselection  selector   specified by a choice of $\{j_1, j_2, \cdots j_n\}$.
 The component of $|\Phi\rangle$ in such a sector lies in a space of dimension 
 \be
 \label{valdd}
 d=\sum_{s=\bar{s}(1-\delta_2)}^{s=\bar{s}(1+\delta_2)}\sum_{t=\bar{t}(1-\delta_3)}^{t=\bar{t}(1+\delta_3)} {m \choose s} {n-m\choose t},
 \ee
 where $m$ is given by eq.(\ref{valmb}). Importantly,  $d$ is the same for all superselection sectors with the same value of $m$. 
 
 As a result, one can  consider the set of all superselection sectors which correspond to the same value of $m$ together.  Alice can   use two sets of $q_m$ qubits which are initially unentangled and in the state
 \be
  \label{inqubit}
  |000\cdots 00\rangle \otimes |000\cdots 00\rangle.
  \ee
and extract the entanglement in all these  superselection sectors. The number of qubits needed are 
 \be
 \label{valqm}
 q_m = S_q(m) (1+ O(1/\sqrt{n}))
 \ee
 where 
 \be
 \label{defhm}
 S_q(m)=-m[\alpha_{1,1}\log\alpha_{1,1}+\alpha_{0,1}log\alpha_{0,1}]-(n-m)[\alpha_{1,0}\log\alpha_{1,0}+\alpha_{0,0}\log\alpha_{0,0}]
 \ee
  This follows from noting that  $d$ defined in eq.(\ref{valdd}) takes the value
  \be
  \label{bondd}
  d=2^{ S_q(m) (1+ O(1/\sqrt{n}))}.
  \ee
      
  To extract the entanglement in these  superselection sectors Alice  swaps the state of the first set of 
   $q_m$ qubits  with that of   $A'$  in each  superselection sector,   and then the state of the  second set of qubits with $C$ also in the same  sector. 
  The second set of qubits is then  teleported to Bob using Bell pairs, following which  Bob then swaps its state with that of B also in the same  superselection sector. 
  Finally,  Alice swaps the state of the first set of qubits  with that of $A$ also in the same  superselection sector.   
  The number of Bell pairs used up in the process is $q_m$. 
  
  Actually, the last two swaps need to be defined more precisely as follows. Bob takes the state of the second set of qubits obtained after teleportation and swaps it with states in a $d$ dimensional subspace of $B$ obtained by taking the quantum numbers $s, t$ defined in eq.(\ref{valbars}), eq.(\ref{valbart}), now referring to the system $B$, in the range, $s\in\{\bar{s}(1-\delta_2), \bar{s}(1+\delta_2)\},
   t \in \{\bar{t}(1-\delta_3), \bar{t}(1+\delta_3)\}$. Similarly for the last swap  transformation that Alice carries out. 
   Also, we note that by adjusting $\delta_2, \delta_3$ (while keeping them to be $O(1/\sqrt{n})$) we can ensure that $q_m$ is an integer.
   
 This procedure can be extended for all values of 
 \be
 \label{rm}
 m\in \{\bar{m}(1-\delta), \bar{m}(1+\delta)\}
 \ee
  in the sum, eq.(\ref{wfproj}), as follows. The number of values $m$ takes
 is 
 \be
 \label{rmaa}
 \Delta m = 2^{\Delta q},
 \ee
 where $\Delta q \sim O(\log \sqrt{n})$. This follows  from eq. (\ref{valma}) and eq. (\ref{valdelta}).  One can therefore keep track of which value of $m$ one is working with by adding an additional $\Delta q$ number of pairs of qubits (this number can  also be made  an integer by adjusting $\delta$). 
 
  For this purpose we   start with 
  \be
  \label{refqt}
  q_T= \Delta q + q_{max}
   \ee number of pairs of qubits, where $q_{max}$ is the maximum value $q(m)$ takes
 as $m$ ranges over the set in eq.(\ref{rm}). It follows from eq.(\ref{valma}), eq.(\ref{valqm}) that 
  \be
 \label{req}
 q_{max}= n S_q (1+ O(1/\sqrt{n})).
 \ee
 
 Each of the $\Delta q$ pairs can be  taken in  a Bell pair state. As a result the state of all of these pairs is a symmetric superposition of $\Delta m$ states, with one set of 
 $\Delta q$ qubits being completely correlated with the other set. The remaining $q_{max}$ pairs are in the states, eq.(\ref{inqubit}). The $q_T$ pairs can then be divided into two sets,
 with each set containing half of the $\Delta q$ Bell pairs and one set of the $q_{max}$ pairs. 
 
 Depending on the state of the   $\Delta q$ qubits in the first set
   Alice  then carries out a swap between the state of $A'$  in the set of  superselection sectors with the corresponding value of $m$ and the $q_{max}$ qubits in the first  set. And she carries out a swap between $C$ for the same value of $m$ and the second set of $q_{max}$ qubits. 
   More correctly, Alice carries out a swap between the state of $A'$ with the appropriate value of  $m$   and a subset of the state of the $q_{max}$ qubits of dimension $q(m)$, eq.(\ref{valqm}), and similarly for $C$. 
   The second set of $q_T$ qubits is then teleported to Bob using Bell pairs, and then swapped with $B$. Once again this is done by relating the values of the first $\Delta q$ qubits to the value of $m$ in the state in possession of Bob, eq.(\ref{strpsi}). 
   And the first set of $q_T$ qubits is then swapped similarly with $A$.

   The total number of Bell pairs consumed in this process is 
 \be
 \label{defkp}
 k'=q_T = n S_q + O(\sqrt{n}) + O\left(\log \sqrt{n}\right).
 \ee

   
 The result of these operations is  that $|\psi^{in}\rangle$, eq.(\ref{strpsi}),  turns into the state $|\Phi\rangle$ which is now shared between $A$ and $B$. From our discussion above it follows that this 
  final state  is  a good copy of 
  $|\psi\rangle^{\otimes n}$ with fidelity $1-\epsilon$. 
 By taking $n \rightarrow \infty$ we can take $\epsilon \rightarrow 0$. 
 
 The number of Bell pairs consumed in this limit becomes
 \be
 \label{totalcon}
 \lim_{n\rightarrow \infty} {k'\over n}=S_q
 \ee
 which agrees with eq.(\ref{vaptp}) above.

 \subsection{Concluding Comments on the Toy Model}
 In the usual discussion of distillation and dilution, without superselection sectors, the end point of distillation is an unentangled state, and so is the starting point of the dilution process. One also finds that the maximum number of Bell pairs produced $k$ during distillation and the minimum number  $k'$ used during dilution satisfy the condition
 \be
 \label{condaa}
 k<k'
 \ee
 with 
 \be
 \label{condba}
 \lim_{n\rightarrow \infty} {k\over n} = \lim_{n\rightarrow \infty} {k'\over n} = S_{EE}
 \ee
 where $S_{EE}$ is the entanglement entropy. 
 This allows one to conclude that the protocols one has identified are optimal. For if one could have carried out the dilution process while consuming fewer Bell pairs,
  then combining that with the known protocol for distillation would lead to a process where one began and ended with the same unentangled state but produced some 
  net number of Bell pairs. This would be a contradiction since only LOCC were involved. 
  
  In the discussion above with superselection sectors  we have seen that $k$ and $k'$ do meet the conditions,\footnote{While we did not elaborate on eq.(\ref{condaa}) above, it essentially follows from the inequality eq.(\ref{condpa}) and eq.(\ref{defkp}).} eq.(\ref{condaa}) and eq.(\ref{condba}). However the end point of distillation given in eq.(\ref{psifinal}) and the state we 
  needed to begin the dilution process, eq.(\ref{strpsi}),  are not the same.
  Thus one might wonder if one has identified the optimal protocols in this case. Perhaps one can improve them and extract more Bell pairs than given in eq.(\ref{condba})?

  We now argue  that this is not possible and the limit $S_q$ identified above is indeed the best one can do. 
  The point is the following. We had already argued in section \ref{toymodel} that using only local gauge-invariant operators  we should not be able  to extract the full entanglement in the Bipartite state. A residual entanglement given by the $S_c$ term in eq.(\ref{t1}) must remain unextracted from the Bipartite state. 
  This expectation is in fact borne out in the two protocols above. It is easy to see that the entanglement per copy in the residual state left over after distillation eq.(\ref{psifinal}) is indeed $S_c$,
  as is the entanglement in the starting state, eq.(\ref{strpsi}), of dilution. 
  
  Thus, the  beginning state of distillation and the end point of dilution in our protocols, while not being the same, do achieve the minimum bound on the residual entanglement that cannot be extracted. It then follows that the maximum entanglement that could have been extracted in Bell pairs is given by $S_q$, the second term in eq.(\ref{enttoy}), eq.(\ref{defha}),  which is indeed met by the protocols above. 
  
  The argument about $S_q$ being the maximum entanglement that can be extracted can be seen as follows. Suppose we could have extracted more entanglement than $S_q$. Then, at the end of the distillation process, tasteless as it  may seem,  working in the extended Hilbert space where we allow non-gauge-invariant operators. we could  also extract the entanglement from the residual state which equals $S_c$. This would result in  a total extracted entanglement which is bigger that $S_c+S_q$, eq.(\ref{enttoy}), thereby leading to a  contradiction. 
  
  \subsection{Comments on Gauge Theories}
    Let us end by  connecting our discussion of the toy model above with that for  gauge theories. As was mentioned in section \ref{toymodel} above, the $S_c$ term, eq.(\ref{t1}), is analogous to the first term in eq.(\ref{defsnt2}) in the Abelian case and the first two terms in eq.(\ref{feeta}) for the Non-Abelian case. We have seen in the toy model that the entanglement corresponding to the $S_c$ term cannot be extracted during distillation or dilution. This suggests that the corresponding  terms  also cannot be extracted in the Abelian and Non-Abelian cases respectively. 
    It follows then that the maximum entanglement one can extract, in the asymptotic sense where we work with $n$ copies and take $n \rightarrow \infty$, is given by the second term in eq.(\ref{defsnt2}),
    \be
    \label{maxea}
    S_{q, A}= - \sum_{\bf{k}} p_{\bf{k}} \Tr_{{\cal H}_{in}^{\bf{k}}} {\tilde \rho}_{in}^{\bf{k}} \log {\tilde \rho}_{in}^{\bf{k}},
    \ee
     for the Abelian case and by the third term in  eq.(\ref{feeta}),  
     \be
     \label{maxena}
     S_{q, NA}=- \sum_{\bf{k}} p_{\mathbf{k}} \Tr_{{\hat {\cal H}}^{\bf{k}}_{in}}{\bar \rho}_{in}^{\bf{k}} \log({\bar\rho}_{in}^{\bf{k}}),
\ee
     for the Non-Abelian case. 
    This conclusion is in fact true and follows quite straightforwardly  from our discussion of the toy model as we will now see. 
    Eq. (\ref{maxea})  and eq.(\ref{maxena})  are some of the main conclusions of this paper.  For a more complete definition of the various symbols appearing in these equations we refer the reader to  section \ref{definition} especially  the discussion around eq.(\ref{defsnt2}) and eq.(\ref{feeta}).

    The toy model can in fact be immediately generalised to apply to an Abelian gauge theory.  As discussed in  section \ref{elcentre},
    the Hilbert space of gauge-invariant states ${\cal H}_{ginv}$ in the Abelian case can be expressed as a sum over tensor products, eq.(\ref{wras}),  where the sum is over different superselection sections labelled by $\bf{k}$.
    

    In the toy model the second and third qubits, in  the wave function, eq.(\ref{fqs}) took  the same values and could not be changed by allowed local operators.  These qubits then encode for  the superselection  sectors in the model. To extend the toy model for  the Abelian   case we simply add   more qubits in both $A$ and $B$, which are enough in number to encode for the full superselection vector ${\bf{k}}$. The  constraint of gauge-invariance is that these qubits must take the same values in $A$ and $B$. Also we increase the dimension of the Hilbert spaces spanned by the first and fourth qubits in eq.(\ref{fqs}) to equal the dimension of ${\cal H}_{ginv, in}^{\bf{k}}, {\cal H}_{ginv,out}^{\bf{k}}$ respectively. Gauge-invariant operators acting on inside  links allow  states in ${\cal H}_{ginv, in}^{\bf{k}}$ to be transformed from one to another in general, and similarly for ${\cal H}_{ginv,out}^{\bf{k}}$. Our discussion of the toy model then applies in a straightforward way to the Abelian case and shows that the maximum entanglement which can be extracted is given by eq.(\ref{maxea}). The residual entanglement left over after distillation, or which must be present at the start of dilution, is given by the first term in eq.(\ref{defsnt2}). 
    
    The extension to the Non-Abelian case is also straightforward. Recall once again, that in the context of the electric centre definition in section \ref{elcentre}, we found that the space of gauge invariant states could be written as a sum over tensor products, with the sum being over   different superselection sectors, eq.(\ref{ginvs}), for the Non-Abelian case as well. The only difference here is that the labels of each superselection sector specify the irreducible representation of the Non-Abelian gauge group under which the total ingoing (or outgoing) flux at each boundary vertex transforms. The extension required for the toy model to describe this case is then also  very similar to the Abelian case of the previous paragraph. The second and third qubit in the toy model are extended to describe the superselection sectors in the Non-Abelian case, and the dimension of the spaces spanned by the first and fourth qubits are extended to span the spaces   ${\cal H}_{ginv, in}^{\bf{k}}$, 
    ${\cal H}_{ginv,out}^{\bf{k}}$, which were defined in section \ref{ecna} for the Non-Abelian case. This leads to the conclusion that the maximum entanglement which can be extracted in the Non-Abelian case is given by the second term in eq.(\ref{entcc}), which is the third term in eq.(\ref{feeta}) and in agreement with eq.(\ref{maxena}).     
    
    The residual entanglement which cannot be extracted depends on the definition which is used. For the electric centre definition this is given by the first term in eq.(\ref{entcc}), which is also the first term in eq.(\ref{feeta}). 
    For the extended Hilbert space definition the residual entanglement is given by the first two terms in eq.(\ref{feeta}) as was mentioned above. 
    
    The extension of the toy model for the Non-Abelian case as described above actually gives the description of the state in ${\cal H}_{ginv}$  which was discussed 
      in section \ref{ecna}. 
   The model can be further extended to directly give  the description of a state in the extended Hilbert space, ${\cal H}$.
   Since ${\cal H}_{ginv,in}^{\bf{k}}, {\cal H}^{\bf{k}}_{ginv,out}$, are already identified with ${\hat{\cal H}}^{\bf{k}}_{ginv,in}$, ${\hat{\cal H}}^{\bf{k}}_{ginv,out}$, respectively, eq.(\ref{ida}), eq.(\ref{idb}), one needs to augment the  Hilbert spaces spanned by the second and third qubits of the toy model  further by also adding the spaces ${\cal H}^1\otimes \cdots \otimes {\cal H}^n$, and ${\cal H}^{'1}\otimes \cdots \otimes {\cal H}^{'n}$ respectively to them for this purpose.
   As was mentioned in the discussion around eq.(\ref{depf}), given a state in ${\cal H}_{ginv,in}^{\mathbf{k}}\otimes {\cal H}_{ginv,out}^{\mathbf{k}}$ we can construct a unique state in ${\cal H}^{\bf{k}}$ from it. Summing over the different  superselection sectors ${\bf{k}}$ then gives the state in ${\cal H}$. In this way, after the additional augmentation the toy model  directly describes states in the extended Hilbert space. It is then straightforward to see that the analysis above when carried out  in this augmented case  leads to the conclusion that the  entropy which can be extracted is given by eq.(\ref{maxena}) and the residual entanglement entropy which cannot be extracted is  the first two terms in eq.(\ref{feeta}).


\section{Degeneracy of the Non-Abelian Toric Code} \label{sec:degeneracy}
    Since, we as saw above, the full   entanglement entropy given by the   extended Hilbert space definition  cannot be extracted in distillation or dilution, one might ask whether an alternate definition given by the entanglement which can be extracted in these processes, by the most optimal protocols, is in fact more appropriate. 
    
    Here, we will examine a class of Non-Abelian toric codes \cite{Kitaev1997} and find that the extended Hilbert space definition has  the attractive feature of  leading to the correct  topological entanglement entropy, \cite{Kitaev2005,Levin2005}. It will turn out in these models that the extra terms which arise in the extended Hilbert space definition (the first two terms in eq.(\ref{feeta})) make an important contribution to the full answer. An alternate definition,  corresponding to the number of Bell pairs which can be extracted, will therefore not lead to the correct topological entropy.


 In fact, this is already true in the Abelian case, where the full entanglement, eq.(\ref{defsnt2}) is needed and the second  term alone is not good enough, \cite{Casini2013}. 
 Here we will discuss the Non-Abelian case to also bring out the significance of the second term in eq.(\ref{feeta})  which arises only in this context.  
  
 The fact that the extended Hilbert space definition gives the correct answer to the topological entanglement entropy for gauge theories  based on  toric codes is actually   more or less guaranteed to be true from their very construction. In these models gauge-invariance is only imposed ``weakly" to begin with,  by a term in the Hamiltonian which makes it energetically unfavourable to break this symmetry. While the  resulting ground state is gauge-invariant the excited states are therefore not, in general. 
 And the correct entanglement entropy is obviously given by  working in the extended Hilbert space.
 The gauge theory is recovered in the limit in which the electric coupling $J_e \rightarrow \infty$ (this coupling is defined in eq.(\ref{htc}) below). It then follows that the natural definition of entanglement in the gauge theory should also be obtained by the extended Hilbert space definition in this limit. Moreover, since the topological entanglement entropy is by construction invariant under smooth deformations of the Hamiltonian, it should not change as this limit is taken (provided correlation lengths continue to be finite, as will be the case here). Thus, the extended Hilbert space definition should lead to the correct topological entanglement entropy in the gauge theory too. 
 
 The purpose of carrying out the explicit calculations here is to illustrate that the extra terms present in the extended Hilbert space definition, do contribute in an essential way in the answer for the topological entanglement entropy. 
 
 We  work in this section with   toric codes   based on  a discrete non-Abelian symmetry group $G$. 
 The dimension of this group will be denoted by $|G|$, and its irreducible representations by $r_i, i= 1, \cdots n$. 
 A concrete example to keep in mind is $S_3$, with dimension $6$ and  three irreducible representations. 
 
 We work in $2$ spatial dimensions. The degrees of freedom live on links $L_{ij}$. For each link these degrees  lie in a   Hilbert space ${\cal H}_{ij}$ of dimension $|G|$.
Suppressing the link indices for the moment, a basis of the Hilbert space at each link is given by the orthonormal vectors, $\{|g\rangle :   g\in G\}$. 
The corresponding basis elements for ${\cal H}_{ij}$ are denoted by $\{|g\rangle_{ij}\}$. A basis for the full Hilbert space ${\cal H}=\otimes_{ij} H_{ij}$ is obtained by taking the tensor product $\otimes |g\rangle_{ij}$ over all the links. 
Note that each link $L_{ij}$ has an orientation. If this link is in the state $|g\rangle_{ij}$ then its oppositely oriented counterpart $L_{ji}$,   would be in the state $|g^{-1}\rangle_{ji}$.

There is a  left and right action of the group which can be defined by operators ${\cal L}_{+, ij}^{h}, {\cal L}_{-,ij}^ h$  on each link as follows:
\be
\label{lact}
{\cal L}_{+,ij}^h|g\rangle_{ij}=|h\cdot g\rangle_{ij}
\ee
\be
\label{Pract}
{\cal L}_{-,ij}^ h|g\rangle_{ij}=|g \cdot h^{-1}\rangle_{ij}
\ee
for any $h \in G$. 
This is analogous to the action of the operators, ${\hat J}^a, {\hat {\cal J}}^a$ in the $SU(2)$ case, eq.(\ref{lact}), eq.(\ref{ract}).

We also define projectors, $T^h_{+,ij}, T^h_{-,ij}$, by 
\be
\label{actproj}
T_{+,ij}^h|g\rangle=\delta_{h,g}|g\rangle_{ij}, \ \ T_{-,ij}^h|g\rangle=\delta_{g^{-1},h}|g\rangle_{ij}
\ee
for any $h\in G$. 

The Hamiltonian is given by the sum of two terms:
\be
\label{htc}
H= -J_e \sum_{i}A_i -J_m \sum_{p} B(p)
\ee
with $i$ denoting all vertices, $V_i$, in the lattice and $p$ denoting the sum over all elementary plaquettes. The couplings,   $J_m, J_e > 0$. 

The operator $A_i$ is defined at each vertex $V_i$  as follows. We take all links $L_{ij}$  to be oriented outward from the vertex, and take,
\be
\label{defaia}
A_i=\sum_{g} \prod_j {\cal L}^{g}_{+,ij}
\ee
where ${\cal L}^{g}_{+ ,ij}$, defined in eq. (\ref{lact}), acts on the state of the $L_{ij}$ link. The sum  in eq.(\ref{defaia}) is over all elements of $G$. 
The product in eq.(\ref{defaia})  takes values  over  all links emanating from $V_i$. Under a gauge transformation, parametrised by group element $h$, at vertex $V_i$, the state $|g\rangle_{ij}$ of the Link $L_{ij}$ which emanates from $i$ and goes to $j$
transforms as $|g\rangle_{ij}\rightarrow \mathcal{L}_{+,ij}^h|g\rangle_{ij}$.   Thus we see that the sum  in $A_i$ corresponds to a  sum over all  gauge transformations at vertex $V_i$ and the first term in $H$ is then a sum over all gauge transformations for all vertices.  
  
The operator $B(p)$ is defined as follows. Consider the elementary plaquette $p$ and orient the links in the plaquette to run counter clockwise. There are four such
 links which we label as $1,2,3,4$. 
Then 
\be
\label{defbp}
B(p)=\sum_{h_1,h_2,h_3,h_4} T_{+, 1} ^{h_1}T_{+, 2}^{h_2} T_{+,3}^{h_3} T_{+,4}^{h_4},
\ee
where the sum is over all values of group elements $h_i,h_2,h_3,h_4,$ subject to the condition $h_1 h_2 h_3 h_4=1$. 
 Loosely speaking the  operator $B(p)$  projects onto states where  the magnetic flux around  the plaquette is identity.
 
 It is easy to see that the two terms in $H$ commute with each other. This makes it easy to find the ground state, $|s\rangle$. 
 This state is unique  on the plane, $\mathbb{R}^2$.  It  is gauge-invariant  and satisfies the conditions,  
 \be
 \label{condags}
 A_i |s\rangle=|s\rangle,
 \ee
  at each vertex,  and  
  \be
  \label{cndpa}
  B(p) |s\rangle=|s\rangle
  \ee
  for each plaquette. 
  The ground state  is given by the symmetric sum of all states   for which   the magnetic flux around each plaquette takes the value identity.
  We will be interested in the topological entanglement entropy of $|s\rangle$. 

  \begin{figure}[h]
    \centering
    \includegraphics[width=50mm]{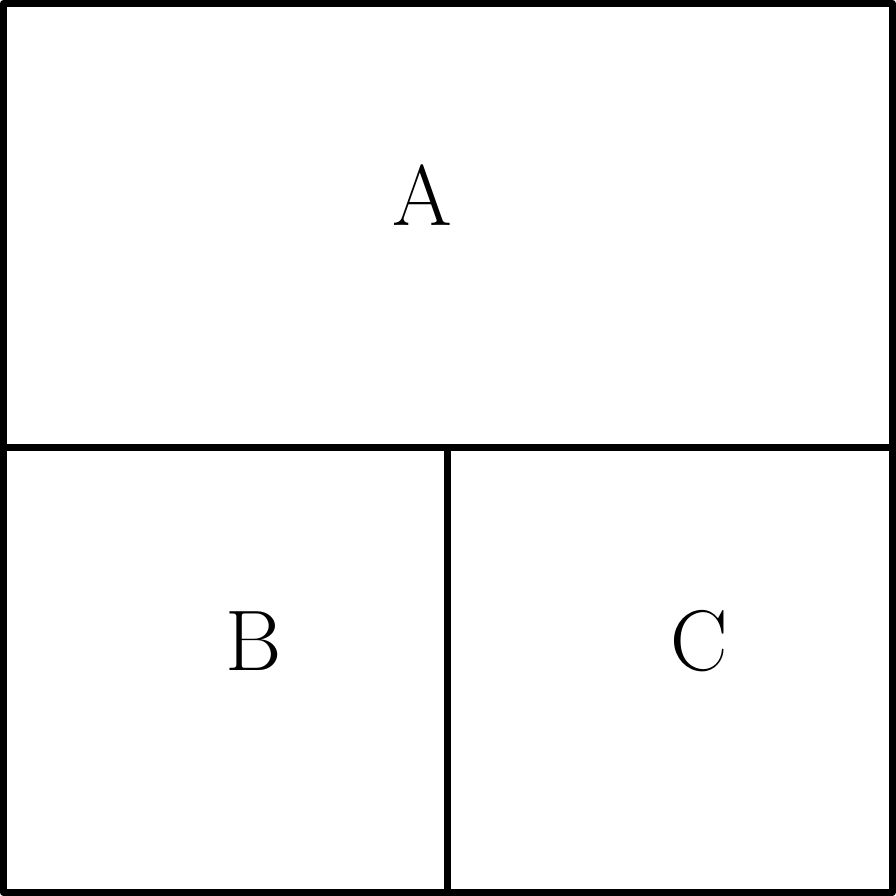}
    \caption{The topological entanglement entropy can be calculated form three regions of this sort, by the formula, eq. (\ref{top}).}
    \label{fig:topeeregions}
  \end{figure}
  
  Consider three regions $A, B, C$ shown in Figure \ref{fig:topeeregions}. 
  Let $S_A$ be the entanglement entropy of $A$ with the rest of the system, $S_{A\cup B}$ the entanglement of $A\cup B$ with the rest, etc.
  Then the topological entanglement is defined as 
  \be
  \label{top}
  S_{top}=S_A+S_B+S_C-S_{A\cup B}-S_{B\cup C}-S_{C\cup A}+S_{A\cup B\cup C}.
  \ee
  
  To compute it  we will first consider a single region  which is the set of links on and inside a rectangle as shown in Fig \ref{fig:generalregion} and show how its entanglement is computed.
  The result for $S_{top}$ will then follow. 
  
  For the rectangular region let there be $n$  boundary links taking values, $1, 2, \cdots n$, ordered in an anticlockwise fashion. And let us consider a configuration where the
  degree of freedom on each of these  boundary links take values,
  $\{h_1, h_2, \cdots h_n\}$, meeting the condition, 
  \be
  \label{consta}
  h_1 h_2 \cdots h_n=1
  \ee
  as required for state $|s\rangle$. Due to the constraint, eq.(\ref{consta}), only $n-1$ of the link variables  are independent resulting in   $|G|^{n-1}$ such independent configurations. 
  Now since the magnetic flux on all plaquettes is completely determined in the state $|s\rangle$, eq.(\ref{cndpa}), it can be shown that the resulting state which contributes to the ground state $|s\rangle$ with the choice $\{h_1, h_2, \cdots h_n\}$ has the form,
    \be
  \label{resstate} 
  |h_1h_2h_3 \cdots h_n\rangle\otimes |\psi_{\{h_i\}, out}\rangle\otimes |\psi_{\{h_i\},in}\rangle.
  \ee
  Here $|\psi_{\{h_i\},out}\rangle, |\psi_{\{h_i\},in}\rangle$, are pure states defined on the Hilbert spaces of the outside and inside links, and $|h_1h_2h_3 \cdots h_n\rangle$ is a shorthand for the state $ |h_1\rangle\otimes|h_2\rangle\otimes  \cdots \otimes |h_n\rangle$ of the boundary links.   Further, the states $|\psi_{\{h_i\},out}\rangle$, in eq.(\ref{resstate}), for distinct choices of the boundary links, $\{h_1, h_2, \cdots h_n\}$,
   are orthogonal. Similarly, the states $|\psi_{\{h_i\},in}\rangle$ are also orthogonal.\footnote{Eq.(\ref{resstate}) can be seen, for example, by   fixing  gauge at all vertices, other than boundary 
   vertices. Such a gauge fixing does not change the entanglement entropy.} 
   
   The state $|s\rangle$ can then  be written down as the sum of all such terms,
   \be
   \label{defstate}
   |s\rangle={1\over |G|^{n-1\over 2}}\sum_{\{h_i\}} |h_1h_2h_3 \cdots h_n\rangle\otimes |\psi_{\{h_i\}, out}\rangle\otimes |\psi_{\{h_i\},in}\rangle,
   \ee
   where the sum is over all values of $\{h_1, h_2, \cdots, h_n\}$ subject to the condition eq.(\ref{consta}). There are $|G|^{n-1}$ orthogonal states in the sum
   resulting in the pre factor in eq.(\ref{defstate})
   which makes $|s\rangle$ of unit norm.

  Tracing over the outside links then gives the density matrix
  \be
  \label{rhotc}
  \rho_{in}={1\over |G|^{n-1}}\sum_{\{h_i\}} |h_1h_2h_3 \cdots h_n\rangle\otimes |\psi_{\{h_i\}, in}\rangle\langle h_1h_2h_3 \cdots h_n|\otimes \langle\psi_{\{h_i\}, in}|
  \ee

  Since the density matrix is diagonal in the basis, $|h_1 h_2 \cdots h_n\rangle\otimes |\psi_{\{h_i\}, in}\rangle$, in fact a multiple of the identity, the entanglement is easily computed.
  We get,
  \be
  \label{entatc}
  S_{EE}= (n-1)\log(|G|)
  \ee
  due to the $n-1$ independent boundary links. 
  
  The topological entanglement can now be easily calculated for the three regions shown in Fig \ref{fig:topeeregions} and gives
  \be
  \label{topea}
  S_{top}=-\log(|G|)
  \ee
  
  \subsection{The Different Superselection Sectors}
  Let us now go back to the rectangular region in Fig \ref{fig:generalregion} and try to understand our result for the entanglement  in terms of contributions from different  superselection sectors   as defined in section \ref{definition}. The links of interest are those inside and on the boundary of the rectangle, their degrees of freedom lie in ${\cal H}_{in}$. 
  We will refer to the   basis $\{|h_1 h_2 \cdots h_n\rangle\}$ used to describe the states of the boundary links above as the ``magnetic basis'' below. 
  The states $|h\rangle$ of a link are not gauge-invariant under a gauge transformation acting on its end points. This means that the basis $\{|h_1h_2 \cdots h_n\rangle\}$ is  not invariant under boundary gauge transformations. By a boundary gauge transformation we mean a gauge transformation acting at one of  the boundary vertices.

  The basis in which the contribution from different superselection sectors becomes manifest is a different one. It is obtained by decomposing the Hilbert space of the inside links, ${\cal H}_{in}$,  into irreducible representations of the boundary gauge transformations. Although   the basis elements $|h_1h_2 \cdots h_n\rangle$  for the boundary links are  not invariant under these transformations, an important point to note is that  the sector satisfying the condition, eq.(\ref{consta}),  
     for which the density matrix has non-trivial support, is  indeed invariant.  
   This sector can therefore by itself  be  decomposed into irreducible representations of the boundary gauge transformations. Notice also, as is clear  from eq.(\ref{defstate}), eq.(\ref{rhotc}), 
   that in this sector the density matrix has a very simple form, namely being a multiple of the identity. 
  It therefore must continue to have this form even when we decompose the sector into the irreducible representations of the boundary gauge transformations.

   Our job of expressing the entanglement in terms of different superselection sectors is therefore quite easy. We have simply to figure out which irreducible representations of the boundary gauge transformations occur  after we carry out this basis change.

  A superselection sector is specified by the vector ${\bf k}= \{r_1, r_2,\cdots ,r_n\}$. States of ${\cal H}_{in}$ lying in this sector transform as the $r_i$ representation under  a boundary gauge transformation at the vertex $V_i$. To see why this specifies a superselection sector, we note that Gauss' law then tells us that the outside links (oriented suitably) leaving $V_i$ must transform 
  as the  conjugate representation, ${\bar r}_i$. Now, the state of the outside   links cannot be changed by operators acting on the inside links alone. Thus the superselection sectors cannot be changed by gauge-invariant operators acting only on the inside region. 
  
 It is easy to see that  the  different representations  $r_1, r_2, \cdots, r_n$  which enter in ${\bf k }$ are not all  independent. The constraint eq.(\ref{consta}) satisfied by the boundary links leads to a  condition amongst them. For given values $r_1, r_2, \cdots r_{n-1}$ of the representations under the  gauge transformations at the first $n-1$ boundary vertices one finds that the only values for $r_n$ which appear are those which can combine with the others to give a singlet of $G$. For the state $|s\rangle$ we started with, this is in fact the  only constraint. Since the state is a symmetric sum over all links satisfying the condition eq.(\ref{consta}), all values of $\{r_1, r_2, \cdots r_n\}$ must appear, subject to this one constraint,  when we decompose the sector, eq.(\ref{consta}), into irreducible representations. While we omit some of the steps here, this result can be obtained by  starting with the basis $|h_1, h_2, \cdots h_n \rangle$ for the boundary links and explicitly changing to a new basis which transforms as an irreducible representation under the boundary gauge transformations.
    
  
  
  As a check we note the following. 
  Let us denote the dimension of the $r_i^{th}$ representation of $G$ to be  $d_i$, so that the $r_1$ representation has dimension $d_1$, etc. Then  the dimension of the representation obtained by taking a tensor product of the,   $\{ r_1, r_2, \cdots r_{n}\}$ representations is $ d_1 d_2 \cdots d_n$. 
  For a given set of values of the first $n-1$ labels, $\{r_1, r_2, \cdots, r_{n-1}\}$ we get a subspace of dimension
    \be
  \label{ntdim}
   d_1 d_2 \cdots d_{n-1}.
  \ee
  For the state to be a singlet under global gauge transformations, $r_{n}$ must take the values corresponding to all the irreducible representations that appear in $r_{1} \otimes \cdots \otimes r_{n-1}$. Therefore, for any given set of values $\{r_{1},\cdots r_{n-1}\}$ the set of values $r_{n}$ can take must satisfy
  \begin{equation}
    \sum_{r_{n}} d_{n} = d_{1}\cdots d_{n-1}.
    \label{condrl}
  \end{equation}
  However, a representation may appear in multiple linearly independent ways in the tensor product of $n-1$ representations; this corresponds to the fact that the identity representation appears in multiple linearly independent ways in the tensor product $r_{1} \otimes \cdots \otimes r_{n}$. Since the superselection sectors are completely specified by the representations $\{r_{1},\cdots r_{n}\}$, all the combinations belong to the same superselection sectors.\footnote{The original version of the paper had an error. It  claimed that only one singlets state could appear in each superselection sector. The discussion, with the correct argument given here appeared in \cite{vanAcoleyen2015}.} We will denote the number of combinations as $N_{\mathbf{k}}$ below. In terms of our discussion in section \ref{nonabelian}, the subspace of ${\cal H}_{in}^{\bf{k}}$, the subspace of ${\cal H}_{in}$ corresponding to the superselection sector ${\bf k}$, on which $\rho_{in}^{\mathbf{k}}$ has support is of the form given in eq.(\ref{tprodh}), with ${\cal H}^i, i=1, \cdots n$, having dimension $d_i$, and  ${\hat{\cal{H}}}^{\bf{k}}_{in}$ being replaced by a subspace of dimension $N_{\bf{k}}$.  
   
    Summing over all values of $r_1, \cdots r_{n}$,   with this one constraint then gives the total dimension,
    to be
    \be
    \label{totdim}
    N_T=\sum_{r_1, r_2, \cdots r_n} d_1 d_2 \cdots d_n= \sum_{r_1, r_2, \cdots r_{n-1}} d_1^2 d_2^2 \cdots d_{n-1}^2= |G|^{n-1},
    \ee
    where we have used eq.(\ref{condrl}), and the relation
    \be
    \label{reldg}
    \sum_r d_r^2 = |G|.
    \ee
     We see that $N_T$ is in fact equal to the total  dimension of the space of states which occur in $|s\rangle$, eq.(\ref{defstate}).    
     
     
    It was mentioned above that after the change of basis $\rho_{in}$ must continue to be a multiple of the identity. 
    Form this it follows that the probability to be in any superselection sector, $p_{\bf{k}}$,  is proportional to the  number  of the states  in the superselection sector for which $\rho_{in}$ has non-trivial support, resulting in, 
       \be
       \label{pk}
       p_{\bf{k}} = {(\prod_{i=1}^nd_i ) N_{\bf{k}}\over |G|^{n-1}}.
       \ee
The entanglement entropy is then given by 
\be
\label{entab}
S_{EE}= -\sum_{\bf{k}}p_{\bf{k}}\log p_{\bf{k}}+\sum_{\bf{k}}p_{\bf{k}} \left(\sum_{i=1}^n\log d_i\right) + \sum_{\bf{k}}p_{\bf{k}} \log N_{\bf{k}}.
\ee
The three terms on the RHS correspond to the three terms in  eq.(\ref{feeta}) respectively. In particular, while first two terms cannot be extracted in distillation or dilution, the last term can be can be measured through these processes. It is easy to see from eq.(\ref{totdim}), eq.(\ref{reldg}) that the entanglement in eq.(\ref{entab}) agrees with eq.(\ref{entatc}).


Now that we have verified that the full answer for the entanglement in the ground state of the toric codes considered here gets a significant contribution from  terms which are  not amenable to extraction in distillation and dilution, it also follows that dropping these terms will not give rise to the correct topological entanglement. 
Let us also note that the topological entanglement entropy on general grounds is related to the total quantum dimension, \cite{Kitaev2005,Levin2005}. 
From this general relation and our result, eq.(\ref{topea}) we see that  total quantum dimension, ${\cal D}$, in the models we have been considering here is given by 

\be
\label{valgamma}
{\cal D}=|G|
\ee

It is easy to verify that this is the correct answer in simple examples. E.g. for $S_3$, we get ${\cal D}=6$, which agrees with the value obtained after summing over the quantum dimensions of all particles in the model (see, e.g., \cite{Lahtinen2006}).

\section{Entanglement Entropy  in the Strong-Coupling Expansion} \label{sec:strong-coupling}
 We have seen in the section above on the Non-Abelian toric code that the entanglement in the ground state arises due to the contribution from the first two terms in eq.(\ref{feeta}). 
 Here, we study an $SU(2)$ gauge theory in the strong-coupling expansion. Working to first non-trivial order we find once again that the entanglement arises entirely due to the first two terms in eq.(\ref{feeta}). This calculation has been done earlier \cite{Donnelly2011,Radicevic2015,Chen2015}, but we reproduce it here to emphasise the role of these terms.\footnote{For an interesting discussion of the relation of this answer and that of the previous section to confinement, see \cite{Radicevic2015,Pretko2015}.} 
 
The set up here is closely related to that in section \ref{nonabelian} where we also discussed the $SU(2)$ theory, and we use the same notation. 
We work on a square lattice. The degrees of freedom on each link $L_{ij}$ lie in a Hilbert space ${\cal H}_{ij}$ with states, $|U_{ij}\rangle$.
The Hamiltonian of this model, see \cite{Kogut1975}, after suitably rescaling, is 
\begin{equation}
  H = \frac{g^{2}}{2} \sum_{(ij)} \hat{J}_{ij}^{2} + \sum_{(ijkl)} \frac{4}{g^{2}} \tr (\hat{P}_{\{ijkl\}}  + \hat{P}^{\dagger}_{\{ijkl\}}),   \label{hamsu}
\end{equation}
where 
\be\label{defP}
\hat{P}_{\{ijkl\}} = \hat{U}_{ij} \hat{U}_{jk} \hat{U}_{kl} \hat{U}_{li}
\ee
is the product of all link variables along  an elementary plaquette formed by the links, $L_{ij}, L _{jk}, L_{kl}, L_{li}$. The sum in the first term is over all links $(ij)$, and in the second term over all elementary plaquettes $(ijkl)$. 
The operator $\hat{J}_{ij}^2= \sum_{a=1}^3 ({\hat J}_{ij}^a)^2$ was described in section \ref{nonabelian}, eq.(\ref{lact}). 
Each link appears once in the sum in the first term in eq.(\ref{hamsu}). Since ${\hat J}_{ij}^2={\cal {\hat J}}_{ij}^2$, eq.(\ref{relhatcal}), eq. (\ref{reltwoa}),
  the orientation of the link does not matter. 

$g^2$ in eq.(\ref{hamsu}) is the coupling constant. We work in the strong coupling limit where $g^2 \gg 1$. The Hamiltonian eq.(\ref{hamsu}) can also be written as 
\be
\label{hamtwo}
H=\frac{g^2}{2}\left[H^0+\frac{3}{2}\epsilon H^1\right],
\ee
where 
\begin{eqnarray}
H^0 & =& \sum_{(ij)} \hat{J}_{ij}^{2}, \label{defh0} \\
H^1 & = & \tr ({\hat P_{\{ijkl\}}}  + \hat{P}^{\dagger}_{\{ijkl\}}), \label{defh1} \\
\epsilon & = & \frac{16}{ 3 g^4}.\label{defepsa}
\end{eqnarray}
The prefactor $\frac{g^2}{2}$ can be absorbed in $H$ by a rescaling giving 
\be
\label{resham}
H= H^0+ \frac{3}{2} \epsilon H^1.
\ee
In the strong coupling limit, $\epsilon \ll 1$. We will be interested in computing the ground state of $H$ in perturbation theory in $\epsilon$ and then computing the entanglement of a rectangular region, shown in Fig \ref{fig:generalregion}. We include the links on the boundary in our choice of the inside set of links.   

To leading order the Hamiltonian is given by $H^0$ and the ground state is an eigenstate with  zero angular momentum for all links, i.e.,  $(J_{ij})^2=0$ for all links $L_{ij}$. 
We denote the link $L_{ij}$ in the zero angular momentum state by $|0\rangle_{ij}$ and the full ground state, which is a tensor product, $\otimes_{(ij)}|0\rangle_{ij}$,  by $|0^{(0)}\rangle$.
 
To construct the first order correction to the ground state wave function, we  note that the operator $\tr(\hat{P}_{\{ijkl\}})$   acting on $|0^{(0)}\rangle$
excites the links appearing in the plaquette  to a state which now carries  $1/2$ units of  angular momentum. 
This follows from the   commutation relations, 
\be
\label{commaa}
[{\hat J}_{ij}^a, {\hat U}_{\alpha \beta}]={-\sigma^a_{\alpha \gamma }\over 2}{\hat U}_{\gamma\beta},
\ee
\be
\label{commab}
[{\cal {\hat J}}_{ij}^a, {\hat U}_{\alpha \beta}]={\hat U}_{\alpha \gamma}{\sigma^a_{\gamma\alpha}\over 2},
\ee
where $\sigma^a, a=, 1,2,3$ are the usual Pauli matrices. 
We also note that  ${\cal {\hat J}}_{ij}^a={\hat J}_{ji}^a$, eq.(\ref{relhatcal}).
Each link $L_{ij}$ has two ends and we can  think of ${\hat J}_{ij}^a$ and ${\hat J}_{ji}^a$ as generating an $SU(2)$ transformation at the $i$ and $j$ vertices respectively. 
From the  commutation relations above   we see that the  state ${\hat U}_{ij}|0\rangle_{ij}$ transforms as a spin 1/2 representation under both of these transformations. 
The  trace in $H^1$, eq.(\ref{defh1}), gives rise to a contraction of indices, and as a result the  excitations of    two links ending at a  vertex of the plaquette  combine to give net zero angular momentum  with respect to the $SU(2)$ transformations  at the vertex. 
 This results in a unique state for the four links, which we denote as $|P_{\{ijkl\}}\rangle$  leading to 
\be
\label{restate1}
\tr({\hat P}_{\{ijkl\}})|0\rangle_{\{ijkl\}}=|P_{\{ijkl\}}\rangle,
\ee
where $|0\rangle_{ijkl}$ stands for the tensor product $|0\rangle_{ij}\otimes |0\rangle_{jk}\otimes|0\rangle_{kl}\otimes|0\rangle_{li}$  of the $4$ links  each in the  state with zero angular momentum. 

Acting with $\tr({\hat P}_{\{ijkl \}}^\dagger)$ we get the same state, 
\be
\label{restate2}
\tr({\hat P}^\dagger_{\{ijkl\}})|0\rangle_{ijkl}=|P_{ijkl}\rangle.
\ee

It follows from standard time independent perturbation theory then that the first order state is given by 
\be
\label{fo}
|0^{(1)}\rangle=  \frac{3}{2} \epsilon {\cal P} {1\over E^0-H^0} {\cal P} H^{(1)} |0^{(0)}\rangle,
\ee
where ${\cal P}$ is the projector onto  the subspace of non-zero energies of $H^0$. 
The full state to this order is then 
\be
\label{full state}
|0\rangle={\cal N} \left[ |0^{(0)}\rangle+ \frac{3}{2} \epsilon {\cal P} {1\over E^0-H^0} {\cal P} H^{(1)} |0^{(0)}\rangle \right],
\ee
where ${\cal N}$ is a normalisation to ensure that $|0\rangle$ has unit norm.

It is easy to see that all states excited by a single plaquette insertion of the form eq.(\ref{restate1}), eq.(\ref{restate2}), have an eigenvalue, $3$, with respect to $H^0$, while $E^0=0$.  This gives,
\be
\label{fs2}
|0\rangle={\cal N} \left[|0^{(0)}\rangle-\frac{\epsilon}{2} H^{(1)} |0^{(0)}\rangle\right].
\ee
Now there are three kinds of plaquettes, those which have links lying completely outside  the region of interest, those which have links lying completely inside, and those which have some inside and outside links (note that as per our definition of the region of interest, the links on the boundary of the rectangle in Fig \ref{fig:generalregion} are also in the inside set). 
We will refer to these as the outside,  inside and boundary plaquettes respectively below. 

This gives, 
\be
\label{fs3}
|0\rangle={\cal N}\left[|0^{(0)}\rangle -\epsilon \left(\sum _{out} |P_{\{ijkl\}}\rangle_{out} + \sum_{in} |P_{\{ijkl\}}\rangle_{in}+ 
\sum_{bd}|P_{\{ijkl\}}\rangle_{bd}\right) \right]
\ee
where  the suffixes out,   in and bd stand for  states obtained by acting with the outside, inside and boundary plaquettes respectively. 

Let $n_{in}, n_{out}, n_{bd}$ be the number of in, out and boundary plaquettes respectively, so that 
\be
\label{defN}
N= n_{in}+n_{out}+n_{bd}
\ee
is the total number of plaquettes. 
Also, a simple calculation shows that the norm of each state $|P_{\{ijkl\}}\rangle$ is
\be
\label{normpla}
\langle P_{\{ijkl\}}|P_{\{ijkl\}}\rangle=1.
\ee
Then we have that the normalisation is
\be
{\cal N}={1\over \sqrt {1+ N \epsilon^2}}.
\ee

It is then easy to see that the  density matrix for the inside links that we get, in the extended Hilbert space definition, is
\begin{eqnarray}
\rho_{in} & = & {1\over 1+ N \epsilon^2 }\left[\left\{|0\rangle_{in}-\epsilon \sqrt{n_{in}}\left({1\over \sqrt{n_{in}} } \sum_{in}|P_{\{ijkl\}}\rangle\right)\right\}\left\{\langle0|_{in}-\epsilon \sqrt{n_{in}}\left({1\over \sqrt{n_{in} }}\sum_{in}\langle P_{\{ijkl\}}|\right)\right\} \right. \nonumber \\
&& + \left. n_{out}  \epsilon^2 |0\rangle_{in}\langle0|_{in}+ \epsilon^2 \sum_{bd}\Tr_{{\cal H}_{out}} |P_{\{ijkl\}}\rangle\langle P_{\{ijkl\}}|\right]
\label{rhosu}
\end{eqnarray}
$|0\rangle_{in}$ stands for the state of the inside links obtained by taking the tensor product of the $|0\rangle_{ij}$ states with vanishing angular momentum at each link, and 
the symbols, $\sum_{in}, \sum_{bd}$, stand for a sum over all inside and boundary plaquettes. 
The first term above arises after tracing over a state in which the outside links are unexcited and have  vanishing angular momentum.  
The second term after tracing out  states in which  outside links   lying in one outside plaquette have been excited,   while  the third term arises from states in which a boundary plaquette has been excited. 
 
In terms of different superselection sectors, we see that the contribution in the first line in eq.(\ref{rhosu})  and the first term in the second line   correspond to the superselection sector where the total angular momentum at each boundary vertex vanishes. 
The second  term  on the second line arises  from sectors where the angular momentum  at two adjacent boundary vertices, connected by the boundary plaquette appearing in the sum, does not vanish.  We see from Fig \ref{fig:generalregion}, that for the rectangular region under consideration each boundary plaquette has three links in the outside and one link in the inside (since boundary links are also in the inside set). The state $|P_{\{ijkl\}}\rangle$, involving a boundary plaquette therefore has   three links outside and one link on the boundary of the rectangular region excited with $1/2$ unit of angular momentum. The two boundary vertices with non-zero angular momentum are the ones on which the boundary link of this plaquette ends. 
The angular momentum, $J_T^2$, eq.(\ref{fin}), at  both these vertices is $3/4$ corresponding to $j=1/2$.
For $n$ boundary links we get $n$ different superselection sectors of this type. 

The entanglement entropy can be calculated by adding the contributions from each superselection sector.  The probability to be in a sector where the electric flux in two adjacent boundary vertices is non-vanishing can be obtained by taking the trace of $\rho_{in}$ in this sector. We get 
\be
\label{proba}
p_{(1/2,1/2)}={\cal N}^2 \epsilon^2  \approx \epsilon^{2} .
\ee
There are $n_{bd}$ such sectors. We have used the subscript $(1/2,1/2)$, to remind ourselves that two boundary vertices have $j=1/2$. The precise boundary vertices are not important, since we get the same probability for all sectors of this kind. 

The probability to be in the sector where the angular momentum at each boundary vertex vanishes is given by taking the trace of the first two terms and given by 
\be
\label{probb}
p_0={\cal N}^2 [1+  \epsilon^2 (n_{in}+n_{out})] \approx 1 - \epsilon^{2}  n_{bd}.
\ee
As a check we see that $n_{bd} p_{(1/2,1/2)}+p_0 = 1$ as required. 

Coming back to the sectors which carry non-trivial total angular momentum it is easy to  see that the resulting vector $\bf{k}$, defined in section \ref{nonabelian} has two non-zero values corresponding to the two adjacent boundary vertices with non-zero angular momentum, and these take values $3/4$, eq.(\ref{valki}). The Hilbert space in the sector ${\cal H}_{in}^{\bf{k}}$ is of the form eq.(\ref{tprodh}), where $n=n_{bd}$ and only two of the Hilbert spaces ${\cal H}^1, \cdots {\cal H}^n$ have dimension $2$ while the others have dimension $1$. Also, it is easy to see that the state in $\hat{\mathcal{H}}_{in}^{\mathbf{k}}$ is pure. The density matrix in this sector   $\rho_{in}^{\bf{k}}$ is  also of the form eq.(\ref{rhw}), where ${\hat \rho}^{\bf{k}}_{in}$ is a number.

The resulting contribution from these sectors to $S_{EE}$ then comes only from the first two terms in eq.(\ref{feeta1}) and is 
\be
\label{cent}
\delta_{(1/2,1/2)} S_{EE}=n_{bd} [-p_{(1/2,1/2)}\log p_{(1/2,1/2)}+ p_{(1/2,1/2)} \log(4)].
\ee

For the sector where no angular momentum is excited at the boundary vertices we get a contribution to $S_{EE}$ from the first term in eq.(\ref{feeta1}), and potentially the last term.
 The reduced density matrix is
\begin{equation}
  \frac{1}{1 + (n_{in} + n_{out})  \epsilon^{2}} \left(
   \begin{matrix}
     1 + n_{out}  \epsilon^{2} & \sqrt{n_{in}} \epsilon\\
     \sqrt{n_{in}} \epsilon & n_{in} \epsilon^{2}\\
      \end{matrix}
      \right) 
  \label{redrho}
\end{equation}
in the basis $\left\{ \ket{0}_{in}, \frac{1}{\sqrt{n_{in}}} \sum \ket{P_{i}^{in}} \right\}$.
This matrix has eigenvalues of order $\epsilon^{4}$, and therefore its  contribution up to $O(\epsilon^2)$  is negligible.
The contribution from this sector is then
\be
\label{centa}
\delta_{(0)}S_{EE}=-p_0\log p_0.
\ee
Summing, the total entanglement is 
\be
\label{totent}
S_{EE}=n_{bd} [-p_{1/2,1/2}\log p_{1/2,1/2}+ p_{1/2,1/2} \log(4)]-p_0\log p_0 \approx n_{bd} \epsilon^{2}  \left[ \log \frac{1}{\epsilon^{2}} + 3 \right]
\ee
where we have set $\log 4 = 2$.
From eq.(\ref{proba}) and eq.(\ref{probb}) we see that this is of order $\epsilon^2 \log(\epsilon^2)$. 

A few more comments are in order. First, we see that the contribution up to the first non-trivial order comes entirely from the first two terms in eq.(\ref{feeta1}), as was mentioned above. 
Second, since our leading answer is, up to a log correction, of $O(\epsilon^2)$, the reader may wonder whether we should have calculated the corrected wave function to second order in perturbation theory. The change in normalisation to this order has already been taken into account by the factor ${\cal N}$. It is easy to see that additional corrections are not needed. 
 Such a correction which we denote as $|0^{(2)}\rangle$ would contribute at this order to  the density matrix only through terms of the form $|0^{(0)}\rangle\langle0^{(2)}|+|0^{(2)}\rangle\langle0^{(0)}|$, where the $|0^{(2)}\rangle$ state is orthogonal to $|0^{(0)}\rangle$. As a result, these terms, which  only contribute to the superselection sector with zero angular momentum entering at each boundary vertex, give a contribution of $O(\epsilon^4)$ which can be neglected. 
 
 Finally,  the lattice theory at strong coupling is well known to be confining. This is related to the fact that leading behaviour in eq.(\ref{totent}) is proportional to 
 the number of boundary plaquettes, $n_{bd}$. Since $n_{bd}$ also equals   the number of boundary links, $n$,  the entanglement is proportional  to  the length of the boundary.  It is interesting to compare these results   with the toric code model studied in  section \ref{sec:degeneracy}.
 Confinement, in the strong coupling theory, is accompanied with electric order.  Electric flux tubes cost energy to excite. 
 As a result Gauss' law is implemented at the boundary in a local way, with the electric flux which 
 enters from  a boundary vertex exiting from an adjacent vertex. This results in the leading behaviour being proportional to the number of boundary links and the topological entanglement entropy of the three regions shown in Fig. \ref{fig:topeeregions} vanishing. 
 In  contrast, in the toric code there is a  a global constraint, that the representations $\{r_1, r_2, \cdots, r_n\}$ must be able to combine into a singlets, which is the analogue of a  global Gauss law.  The constraint is a global one since the cost for electric flux tubes to stretch across the inside is not significant. 
 As a result the electric flux on   boundary vertices can take all values subject to one overall constraint. This one global constraint results in the entanglement being  proportional to $n-1$, eq.(\ref{entatc}), leading to  a non-trivial topological entanglement.

  \section{Conclusions} \label{sec:conclusion}
In this paper we  explored the extended Hilbert space definition for the entanglement entropy of gauge theories in some detail. 
The entropy  can be expressed as a sum over different superselection sectors. These sectors correspond to different values for the electric flux, specified in a gauge-invariant way, which can enter the links of interest at each of the boundary vertices. The resulting answer, in each sector, gets a contribution from two kinds of terms. 
The first kind is a sort of classical contribution which arises from the probability to be in this  sector. The second kind is  a quantum contribution due to quantum correlations in the sector. 
In the Non-Abelian case the classical contribution includes an extra term  which arises as follows. In each sector with nonzero electric flux at a boundary vertex, the set of inside and outside links terminating at this vertex must transform in a non-trivial representation of the gauge group. Due to  gauge-invariance these non-trivial representations for the inside and outside links then combine and appear in the wave function only as a singlet of the gauge symmetry. This leads to additional entanglement between the inside and outside links  in the Non-Abelian case since the irreducible representations are of dimension greater than one,
leading to the  extra term in the classical contribution.  The result for the entanglement entropy in  the extended Hilbert space definition is given in eq.(\ref{feeta}). The classical terms are the first two on the RHS in eq.(\ref{feeta}), of these the second term is the one which only arises in the Non-Abelian case. 

We also examined   the processes of entanglement distillation and dilution for gauge theories  in some detail and showed that the number of Bell pairs which can be extracted in these
 processes  are equal, in an asymptotic sense, with a large number of copies of the system, to  only  the quantum type of contribution, i.e. the third term in eq.(\ref{feeta}). 
We argued on general grounds that the quantum contribution was the maximum bound which one could hope to achieve,  since local operations involving gauge-invariant operators cannot change the superselection sectors. 
And we also exhibited protocols for distillation and dilution where the number of extracted or consumed Bell pairs achieved this bound. 
The operational definition in terms of the number of Bell pairs one can extract therefore  gives a smaller value for the entanglement entropy than the extended Hilbert space definition. 

An electric centre definition, following \cite{Casini2013} in the Abelian case, was also presented for   Non-Abelian theories  in the paper. It was  found, interestingly, that this too  differed from  the extended Hilbert space definition. The difference arises only for the Non-Abelian case in  the second term in eq.(\ref{feeta}) which is absent in the electric centre definition. This result is therefore
``midway" between the extended Hilbert space definition and the operational measure described in the previous paragraph: it contains the first and third terms in eq.(\ref{feeta}) but not the second term.

These differences between various definitions  seem bewildering. Which definition should one trust? 

One reason that the  entanglement entropy is useful is that it often provides answers to physical questions.  Since the questions are physical the answers are unambiguous.   For example, the entanglement entropy can be used  to define    the topological entanglement entropy which in turn is often related to the degeneracy of ground states  on a manifold of non-trivial topology or the quantum dimension of excitations in the system. We calculated the topological entanglement entropy  using the extended Hilbert space definition 
for a class of Toric code models based on discrete Non-Abelian gauge groups  here and showed that it gives the correct result for the quantum dimension.  The classical terms make a significant contribution to the full answer in   these models, including an important contribution coming from the second term in eq.(\ref{feeta}) which arises only in the Non-Abelian case. Dropping some or all of these classical contributions would therefore not yield the correct result. 

These investigations suggest to us that the extended Hilbert space definition is indeed a useful one to use.  That it does not agree  with the   operational measure   of the number of Bell pairs that can be extracted from the system may be viewed as  a distinctive feature of gauge theories.  This property is closely tied  to the  fact that non-local excitations corresponding to Wilson loops and electric flux loops  are present in these theories. It seems important to us to study this universal feature of gauge theories further.

There are many other ways in which this work can be extended.  
While we  did not comment on gauge theories with matter explicitly in the paper, it is in fact easy to apply the extended Hilbert space definition for  such systems as well. 
In this case the matter degrees of freedom live on vertices of the lattice, and the system of interest, the ``in'' system,  is then specified by a state on a set of links and vertices. The extended Hilbert space consists of the space obtained by taking the tensor product of link Hilbert spaces for the gauge field degrees of freedom  and vertex Hilbert spaces for matter degrees of freedom. Like in the pure gauge field case, this Hilbert space too  by definition admits  a tensor product structure. Thus tracing over the outside degrees of freedom,  which are in the complement of the inside degrees,   we can compute $\rho_{in}$, eq.(\ref{defrho}),  and then the entanglement entropy, eq.(\ref{defee}).

 One of the properties which makes the extended Hilbert space definition attractive is that it agrees  with a path integral calculation which implements the replica trick. It would be  useful to study the continuum limit of the lattice system we have investigated here  and  analyse the entanglement entropy in this limit. On the lattice it is easy to see that the extended Hilbert space definition is unchanged if one fixes the gauge for vertices securely in the inside or outside. However, gauge fixing the vertices which lie on the boundary, where some links inside and outside terminate, in general changes the result. 
Going to the continuum, starting from this  lattice definition, should help us clarify what kind of gauge fixing is allowed in the continuum too. One expects the results for finite quantities, related to mutual information or topological entanglement entropy for example, should be more universal in the continuum limit.

Finally, 
in $2$ space dimensions it is well known that gauge theories are dual to spin systems. The duality is non-local in space-time. 
It would be useful to understand how  the entanglement entropy transforms under this duality, both in the lattice and  in the continuum. 

We leave these and other related fascinating questions for the future.

\section{Acknowledgements}
We acknowledge discussion with  Sumit Das, Rahul Dandekar, Marina Huerta, Gowtham Raghunath Kurri, Gautam Mandal, Shiraz Minwalla, Pratyush Nath, Pranjal Nayak, Hirosi Ooguri, \DJ or\dj e Radi\v cevi\'c, Sarath Sankar, Ashoke Sen, Tadashi Takayanagi and Spenta Wadia. We especially thank Sudip Ghosh for several discussions and initial collaboration. SPT thanks the organisers of the "International Workshop on Condensed Matter Physics and AdS/CFT", held at IPMU, Japan, from May 25-29, 2015, and  acknowledges the J. C. Bose Fellowship, Department of Science and Technology, Government of India.  We acknowledge  support from  Department of Atomic Energy, Government of India. Most of all, we thank the people of India for  generously supporting research in string theory.

\end{document}